\documentclass[twocolumn,aps,prc,superscriptaddress,showpacs]{revtex4}
\usepackage{amsmath,bm}%
\usepackage{graphicx}%

\begin{document}

\draft
%\preprint{}
\title{Critical Behavior in Light Nuclear Systems:
Experimental Aspects}

\author{Y. G. Ma}
\affiliation{Cyclotron Institute, Texas A\&M University, College
Station, Texas, USA} \affiliation{Shanghai Institute of Applied
Physics, Chinese Academy of Sciences, Shanghai 201800, China}
\author{J. B. Natowitz}
\author{R. Wada}
\author{K. Hagel}
\author{J. Wang}
\affiliation{Cyclotron Institute, Texas A\&M University, College
Station, Texas, USA}
\author{T. Keutgen}
\affiliation{UCL, Louvain-la-Neuve, Belgium}
\author{Z. Majka}
\affiliation{Jagiellonian University, Krakow, Poland}

\author{M. Murray}
\affiliation{University of Kansas, Lawrence KS 66045}

\author{L. Qin}
\author{P. Smith}
\affiliation{Cyclotron Institute, Texas A\&M University, College
Station, Texas, USA}

\author{R. Alfaro}
\affiliation{Instituto de Fisica, UNAM, Mexico City, Mexico}
\author{J. Cibor}
\affiliation{Institute of Nuclear Physics, Krakow, Poland}
\author{M. Cinausero}
\affiliation{INFN Laboratori Nazionali di Legnaro, Legnaro, Italy
}
\author{Y. El Masri}
\affiliation{UCL, Louvain-la-Neuve, Belgium}
\author{D. Fabris}
\affiliation{INFN and Dipartimento di Fisica, Padova, Italy    }
\author{E. Fioretto}
\affiliation{INFN Laboratori Nazionali di Legnaro, Legnaro, Italy
}
\author{A. Keksis}
\affiliation{Cyclotron Institute, Texas A\&M University, College
Station, Texas, USA}

\author{ M. Lunardon}
\affiliation{INFN and Dipartimento di Fisica, Padova, Italy    }
\author{A. Makeev}
\affiliation{Cyclotron Institute, Texas A\&M University, College
Station, Texas, USA}
\author{N. Marie}
\affiliation{LPC, Universit\'e de Caen, Caen, France}
\author{ E. Martin}
\affiliation{Cyclotron Institute, Texas A\&M University, College
Station, Texas, USA}
\author{A. Martinez-Davalos}
\author{A. Menchaca-Rocha}
\affiliation{Instituto de Fisica, UNAM, Mexico City, Mexico}
\author{G. Nebbia}
\affiliation{INFN and Dipartimento di Fisica, Padova, Italy    }
\author{G. Prete}
\affiliation{INFN Laboratori Nazionali di Legnaro, Legnaro, Italy
}

\author{V. Rizzi}
\affiliation{INFN and Dipartimento di Fisica, Padova, Italy    }
\author{A. Ruangma}
\author{D. V. Shetty}
\author{G. Souliotis}
\affiliation{Cyclotron Institute, Texas A\&M University, College
Station, Texas, USA}
\author{P. Staszel}
\affiliation{Jagiellonian University, Krakow, Poland}
\author{M. Veselsky}
\affiliation{Cyclotron Institute, Texas A\&M University, College
Station, Texas, USA}
\author{G. Viesti}
\affiliation{INFN and Dipartimento di Fisica, Padova, Italy    }
\author{E. M. Winchester}
\author{S. J. Yennello}
\affiliation{Cyclotron Institute, Texas A\&M University, College
Station, Texas, USA}

%\email{}

%\address{}
\date{\today}

\begin{abstract}

An extensive experimental survey of the features of the
disassembly of a small quasi-projectile system with $A \sim$ 36,
produced in the reactions of 47 MeV/nucleon $^{40}$Ar + $^{27}$Al,
$^{48}$Ti and $^{58}$Ni, has been carried out.  Nuclei in the
excitation energy range of 1-9 MeV/u have been investigated
employing a new method to reconstruct the quasi-projectile source.
At an excitation energy $\sim$ 5.6 MeV/nucleon many observables
indicate the presence of maximal fluctuations in the de-excitation
processes.  These include the normalized second moments of the
Campi plot  and normalized variances of the distributions of order
parameters such as the atomic number of the heaviest fragment
$Z_{max}$ and the total kinetic energy. The evolution of the
correlation of the atomic number of the heaviest fragment with
that of the second heaviest fragment and a bimodality test are
also consistent with a transition in the same excitation energy
region. The related phase separation parameter, $S_p$, shows a
significant change of slope at the same excitation energy. In the
same region a $\Delta$-scaling analysis for of the heaviest
fragments exhibits a transition to $\Delta$ = 1 scaling which is
predicted to characterize a disordered phase. The fragment
topological structure shows that the rank sorted fragments obey
Zipf's law at  the point of largest fluctuations providing another
indication of a liquid gas phase transition. The Fisher droplet
model critical exponent $\tau$ $\sim$ 2.3 obtained from the charge
distribution at  the same excitation energy is close to the
critical exponent of the liquid gas phase transition universality
class. The caloric curve  for this system shows a monotonic
increase of temperature with excitation energy and no apparent
plateau. The temperature at the point of maximal fluctuations is
$8.3 \pm 0.5$ MeV. Taking this temperature as the critical
temperature and employing the caloric curve information we have
extracted the critical exponents $\beta$, $\gamma$ and $\sigma$
from the data.  Their values are also consistent with the values
of the universality class of the liquid gas phase transition.
Taken together, this body of evidence strongly suggests a phase
change in an equilibrated mesoscopic system at, or extremely close
to, the critical point.

\end{abstract}



\maketitle
\section{introduction}

Probing the liquid gas phase transition of finite nuclei is an
important topic in nuclear physics since it should allow
investigation of   the nuclear equation of state and clarify the
mechanism by which the nucleus disassembles when heated. This
phase transition is expected to occur as the nucleus is heated to
a moderate temperature so that it breaks up on a short time scale
into light particles and  intermediate mass fragments (IMF). Most
efforts to determine the critical point for the expected liquid
gas phase transition in finite nucleonic matter have focused on
examinations of the temperature and excitation energy region
\cite{JBN0,JBN1} where maximal fluctuations in the disassembly of
highly excited nuclei are observed \cite{Campi}. A variety of
signatures have been employed in the identification of this region
\cite{Richert_PhysRep,Dasgupta_01,Bonasera,Chomaz_INPC,Moretto_INPC}
and several publications \cite{Pan,Gulminelli,Elliott} have
reported the observation of apparent critical behavior. Fisher
Droplet Model analysis have been applied to extract critical
parameters \cite{Elliott_PRL02}. The derived parameters are very
close to those observed for liquid-gas phase transitions in
macroscopic systems \cite{Fisher}. Data from the EOS
\cite{Hauger_PRC98} and ISiS \cite{Beaulieu,Bauer} collaborations
have been employed to construct a co-existence curve for nucleonic
matter \cite{Elliott_PRL02}.  Interestingly, the excitation energy
at which the apparent critical behavior is seen is closely
correlated with the entry into the plateau region in the
associated caloric curve \cite{JBN0}. Although implicit in the
Fisher scaling analyses is the assumption that the point of
maximal fluctuations is the critical point of the system
\cite{Elliott_PRL02,Elliott_PRC03}, other theoretical and
experimental information suggest that the disassembly may occur
well away from the critical point
\cite{Norenberg,JBN0,Mekjian,DasGupta,Dorso,Bonasera}. In
addition, recent lattice gas calculations indicate that the Fisher
scaling may be observed at many different densities
\cite{Chomaz-Gulminelli}, raising doubts about previous critical
point  determinations. Further, applications of $\Delta$-scaling
analysis indicate that the observation of power-law mass
distributions \cite{Botet_PRL01,Frankland1}, although necessary,
is not sufficient to identify the true critical point of the
system. We note also that while the role of the long range Coulomb
interaction in determining the transition point has received
considerable theoretical and experimental attention
\cite{Bonche,Song,Richert,EOS-COULOMB,JBN_Preprint02,DasGupta} a
number of questions remain as to the appropriate way to deal with
the complications it introduces.

In this paper we report results of an extensive investigation of
nuclear disassembly in nuclei of $A$ $\sim$ 36, excited to
excitation energies as high as 9 MeV/nucleon. To our knowledge,
this is the smallest system for which such an extensive analysis
has been attempted. An earlier brief report on some aspects of
this work appeared in Physical Review \cite{Ma-nimrod}. While
investigating a smaller system takes us farther from the
thermodynamic limit, several theoretical studies indicate that
phase transition signals should still be observable
\cite{Chomaz,Raduta} in small systems. The choice of a lighter
system for investigation has the advantage of reducing the Coulomb
contributions. Earlier investigations in this mass region have
provided valuable insights into the binary reaction mechanism
\cite{Peter,Ma_PLB97,Mechan,Mechan1,Mechan2,Mechan3},
multifragment emission as a function of excitation energy
\cite{Ref-rec1,Ref-rec2,Ref-rec3,Ref-rec4,Ref-rec5,INDRA,INDRA2},
the emission time-scale and emission sequence of light particles
\cite{Ghetti,Ghetti2,Ghetti3} and collective flow behavior
\cite{Angelique} etc.

Applying a wide range of methods we find that the maximum
fluctuations in the disassembly of A $\sim$ 36  occur at an
excitation energy of 5.6$\pm$0.5 MeV and a temperature of
8.3$\pm$0.5 MeV. At this same point, the critical exponents
describing the fragment distributions are found to be very close
to those of the universality  class of the liquid gas phase
transition.

These observations do not guarantee critical behavior has
been reached, however, in contrast to experimental results for
heavier systems \cite{JBN0} we also find that  the caloric curve
for A $\sim$ 36 does not exhibit a plateau at the point of maximum
fluctuations. Taken together, the observations strongly suggest a
phase change in an equilibrated mesoscopic system at, or extremely
close to, the critical point.

The paper was organized as follows: in Sec. II we describe the
set-up of our experiment; Section III presents a new method for
reconstruction of the quasi-projectile source, QP; Section IV
discusses some general features of the reconstructed QP; Section V
explores the evidence for critical behavior in the disassembly of
the QP. In Section VI, we discuss the caloric curve of the QP In
Section VII we use the scaling theory to derive the critical exponents
of the transition. All those values are found to be consistent
with the universality class of the liquid gas phase transition.
Conclusions are presented in Section VIII.

\section{Experimental Set-up and Event Selection}

Using the TAMU NIMROD (Neutron Ion Multidetector for Reaction
Oriented Dynamics) \cite{Wada_NIMROD} and beams from the TAMU K500
super-conducting cyclotron, we have probed the properties of
excited projectile-like fragments produced in the reactions of 47
MeV/nucleon $^{40}$Ar + $^{27}$Al, $^{48}$Ti and $^{58}$Ni.
Earlier work on the reaction mechanisms of near symmetric
collisions of nuclei in the $20 < A < 64$ mass region at energies
near the Fermi energy have demonstrated the essential binary
nature of such collisions, even at relatively small impact
parameters \cite{Peter}. As a result, they prove to be very useful
in preparing highly excited light nuclei with kinematic properties
which greatly simplify the detection and identification of the
products of their subsequent de-excitation \cite{Steckmeyer}.

The charged particle detector array of NIMROD, which is set inside
a neutron ball, includes 166 individual CsI detectors arranged in
12 rings in polar angles from $\sim$ $3^\circ$  to $\sim$
$170^\circ$. Eight forward rings have the same geometrical design
as the INDRA detector, but have less granularity \cite{Tabacaru}.
The angles, number of segments in each ring and solid angle of
each CsI segment are given in Table I.
\begin{table}
 \caption{NIMROD Charge Particle Array}
 \begin{tabular}{llll}
  \toprule
   Ring & Angle (deg) & No. of Segments & Solid Angle(src)\\
 \colrule

    1 & 4.3 & 12 & 0.96\\
    2 & 6.4 & 12 & 2.67\\
    3 & 9.4 & 12 & 4.26\\
    4 & 12.9 & 12 & 7.99\\
    5 & 18.2 & 12 & 16.1\\
    6 & 24.5 & 24 & 12.7\\
    7 & 32.1 & 12 & 33.6\\
    8 & 40.4 & 24 & 27.6\\
    9 & 61.2 & 16 & 154\\
   10 & 90.0 & 14 & 207.0\\
   11 & 120.0 & 8 & 378.0\\
   12 & 152.5 & 8 & 241.0\\
   \colrule
 \botrule
\end{tabular}
\label{tab1}
\end{table}

In these experiments each forward ring included two
¡°super-telescopes¡± (composed of two Si-Si-CsI detectors) and
three Si-CsI telescopes to identify intermediate mass fragments.
The CsI detectors are Tl doped crystals read by photo-multiplier
tubes. A pulse shape discrimination method using different
responses of fast and slow components of the light output of the
CsI crystals is employed to identify particles \cite{Bernachi}.
In the CsI detectors Hydrogen and Helium isotopes were clearly
identified and Li fragments are also isolated from the heavier
fragments. In the super-telescopes, all isotopes with atomic
number $Z\leq 8$ were clearly identified and in all telescopes
particles were identified in atomic number.

The NIMROD neutron ball, which surrounds the charged particle
array, was used to determine the neutron multiplicities for
selected events. The neutron ball consists of two hemispherical
end caps and a central cylindrical section. The hemispheres are
upstream and downstream of the charged particle array. They are
150 cm in diameter with beam pipe holes in the center. The central
cylindrical sections 1.25m long with an inner hole of 60 cm
diameter and 150 cm outer diameter. It is divided into four
segments in the azimuthal angle direction. Between the hemispheres
and the central section, there are 20 cm air gaps for cables and a
duct for a pumping station. The neutron ball is filled with a
pseudocumene based liquid scintillator mixed with 0.3 weight
percent of Gd salt (Gd 2-ethyl hexanoate). Scintillation from a
thermal neutron captured by Gd is detected by five 5-in phototubes
in each hemisphere and three phototubes in each segment of the
central section.

\begin{figure}
\includegraphics[scale=0.3]{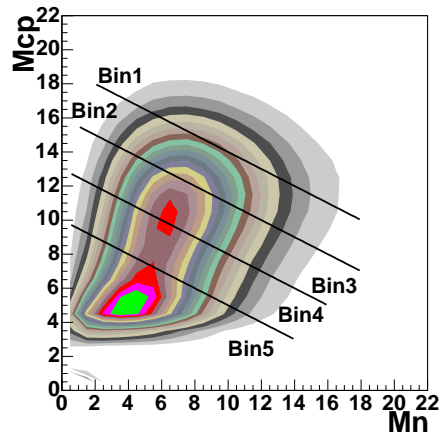}
\caption{\footnotesize  (Color online) Two dimension plot for Mcp
vs Mn as a selector of collision centrality in the $^{40}$Ar +
$^{58}$Ni reaction.} \label{fig_bin}
\end{figure}

\begin{figure}
\vspace{-0.6truein}
\includegraphics[scale=0.3]{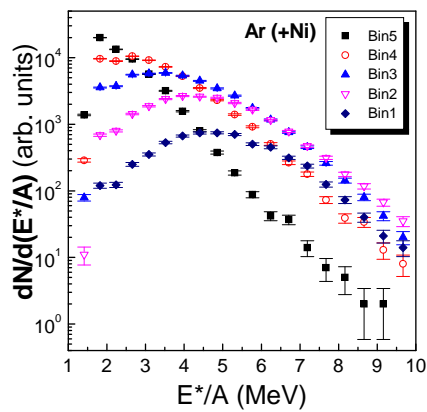}
\vspace{-0.3truein} \caption{\footnotesize  (Color online) The
distribution of excitation energy for different centrality bins
for the QP formed in $^{40}$Ar + $^{58}$Ni.} \label{fig_exc_bin}
\end{figure}

\begin{figure}
\vspace{-0.5truein}
\includegraphics[scale=0.4]{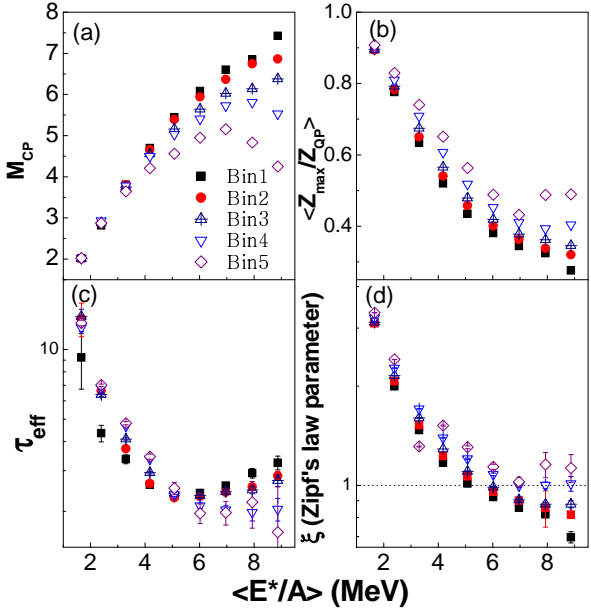}
\vspace{-0.4truein} \caption{\footnotesize  (Color online)
Comparison of some physical quantities as a function of excitation
energy bin in different centrality bins for the QP formed in
$^{40}$Ar + $^{58}$Ni. (a) The total multiplicity of charged
particles; (b) The QP normalized charge number of the heaviest
fragment; (c) The effective Fisher's power-law parameter and (d)
Zipf's law parameter.  The deviations in higher excitation energy
of $Bin5$ and $Bin4$ from $Bin1$ and $Bin2$ can be attributed to
the event mixture in the peripheral collisions.  For detailed
explanation of the physical quantities, see the following
sections. } \label{fig_Bin_comp}
\end{figure}

The correlation of the charged particle multiplicity ($M_{cp}$)
and the neutron multiplicity ($M_n$) was used to sort event
violence. In Fig.~\ref{fig_bin}, lines indicate the event windows
which have been explored.  Roughly speaking , the more violent
collisions correspond to those with  the highest combined neutron
multiplicity  and charged particle multiplicity
($M_{cp}$)($Bin1$). This can be seen in the excitation energy
distribution of the QP in Fig.~\ref{fig_exc_bin} (the
determination of excitation energy will be explained in the
following section).  For $Bin1$ and $Bin2$, the average excitation
is $\sim$ 4 MeV/nucleon and  the $E^*/A$ distribution extends to 9
MeV. Since the goal of the present work was to explore the
disassembly of highly excited QP, we  have used the data from
$Bin1$ and $Bin2$ together in the present work. For the more
peripheral bins, i.e. $Bin4$ and $Bin5$, however, there is
apparent event mixing in the upper range of excitation energy.  In
that case, events coming from a particular excitation energy can
be  distributed over several experimentally reconstructed
excitation energy bins. This judgment is supported by the data
plotted in Figure 3, where we plot, for the QP formed in $^{40}$Ar
+ $^{58}$Ni reactions,  the total multiplicity of charged
particles (Fig.~\ref{fig_Bin_comp}(a)), the QP normalized charge
number of heaviest fragment (Fig.~\ref{fig_Bin_comp}(b)), the
effective Fisher's power-law parameter $\tau_{eff}$
(Fig.~\ref{fig_Bin_comp}(c)) and Zipf's law parameter $\xi$
(Fig.~\ref{fig_Bin_comp}(d)) as a function of the excitation
energy in different centrality bins (for detailed explanations of
the physical quantities plotted, see the following sections). As
seen in the figure, $Bin1$ and $Bin2$ display essentially
identical behavior and the values do not depend the selected
centrality bin. However, data for the peripheral bins ($Bin5$ and
$Bin4$) deviate significantly from the data for $Bin1$ and $Bin2$
in the upper range of the reconstructed excitation energy. These
deviations indicate event mixing in peripheral collisions and
raises questions about  the validity of the excitation energy
determination in the upper excitation energy range for peripheral
collisions. For the intermediate $Bin3$, results are  close to
those of $Bin1$ and $Bin2$. To minimize the effects of possible
event mixing and realize our goal of  exploring  the disassembly
of highly excited QP, we choose only data for $Bin1$ and $Bin2$
and combine those results  for further analysis.

Given the limitation of IMF identification in NIMROD detectors
which do not have Si associated with them, the events with
complete, or near complete QP detection have to be isolated before
the analysis of the QP features can proceed. In  the following
section we describe the techniques of QP reconstruction and event
selection which we have employed .

\section{ A New Method of Quasi-Projectile Reconstruction }

Intermediate energy heavy ion collisions are complicated processes
in which the roles of the mean field and nucleon-nucleon interactions
may both be important. Many reactions manifest the mixed features of
both  the low energy deep inelastic scattering mechanism and a
high energy participant-spectator mechanism.

It is well known that the laboratory  frame kinetic energy spectra
of most  light ejectiles can be reproduced with the assumption of
emission from three different sources: a Quasi- Projectile (QP)
source, an intermediate velocity or Nucleon-Nucleon (NN) source
and a Quasi-Target (QT) source. To better understand origins of
the emitted particles the ideal situation would be to have the
ability to attribute each particle to its source on an
event-by-event basis. However the spectral distributions from the
different sources overlap significantly, making such an
attribution not possible. Previous techniques to reconstruct QP
have included identifying high velocity components of the QP
\cite{Planeta,Sosin} or treating only particles emitted in the
forward hemisphere in the projectile frame and then assuming
identical properties for particles emitted in the backward
hemisphere in order to recreate the QP source
\cite{Peter,Ma_PLB97}. Such a technique is limited in its
application and not suited to situations in which fluctuations are
to be investigated.

For this work we have developed a new method for the assignment of
each light charged particle  (LCP) to an emission source. This is
done with a combination of three source fits and Monte-Carlo
sampling techniques. We first obtain the laboratory energy spectra
for different LCP at different laboratory angles and reproduce
them using the three source fits. In the laboratory frame, the
energy spectra of LCP can be modelled as the overlap of emission
from three independent moving equilibrated sources, {\it i.e.}the
QP, NN and QT sources. For evaporation from the QT source, we take
\cite{Awes}
\begin{equation}
(\frac{d^2N}{dE_{lab} d\Omega_{lab} })_L^{QT}  =  \frac{M_i}{4 \pi
T_s^2} E" \sqrt{E_{lab}/E'} exp(- \frac{E"}{T_s})
\end{equation}
where $E_{lab}$ and $T_s$ are  respectively the laboratory energy
and apparent slope temperature.  $M_i$ are multiplicities. In the
above formula, the Coulomb barrier is considered to be in the QT
source frame, in this case,  $E'$ and $E"$ are defined as
\begin{equation}
E'= E_{lab}-2\sqrt{E_{lab}\frac{1}{2}m_{LCP}v_{s}^2} cos(\theta) +
\frac{1}{2}m_{LCP} v_s^2
\end{equation}
and
\begin{equation}
E" = E'-V_C.
\label{eq_E"}
\end{equation}
where $v_s$ is the magnitude of the source velocity and is taken
along the beam direction.  $\theta$ is the angle between the
source direction and that of  the detected LCP.

For the LCP from QP and NN, we take the Coulomb barrier in the laboratory
frame \cite{Prindle}. For QP, we assume the surface emission form
\begin{equation}
(\frac{d^2N}{dE_{lab} d\Omega_{lab} })_L^{QP}  =  \frac{M_i}{4 \pi T_s^2}
\sqrt{E'E"} exp(-
\frac{E"}{T_s}).
\end{equation}
and for NN, we take the volume emission form,
\begin{equation}
 (\frac{d^2N}{dE_{lab}d\Omega_{lab} })_L^{NN} = \frac{M_i}{2 (\pi
T_s)^{\frac{3}{2}} }
\sqrt{E'} exp(-\frac{E"}{T_s})
\end{equation}
where $E'$ and $E"$ are defined as
\begin{equation}
E' = E_{lab} - V_C,
\end{equation}
and
\begin{equation}
E" = E'-2\sqrt{E'\frac{1}{2}m_{LCP}v_s^2} cos(\theta) +\frac{1}{2}m_{LCP}v_s^2.
\end{equation}

The total energy distribution is the sum over the QP, QT and NN
component.

\begin{figure}
\vspace{-0.2truein}
\includegraphics[scale=0.45]{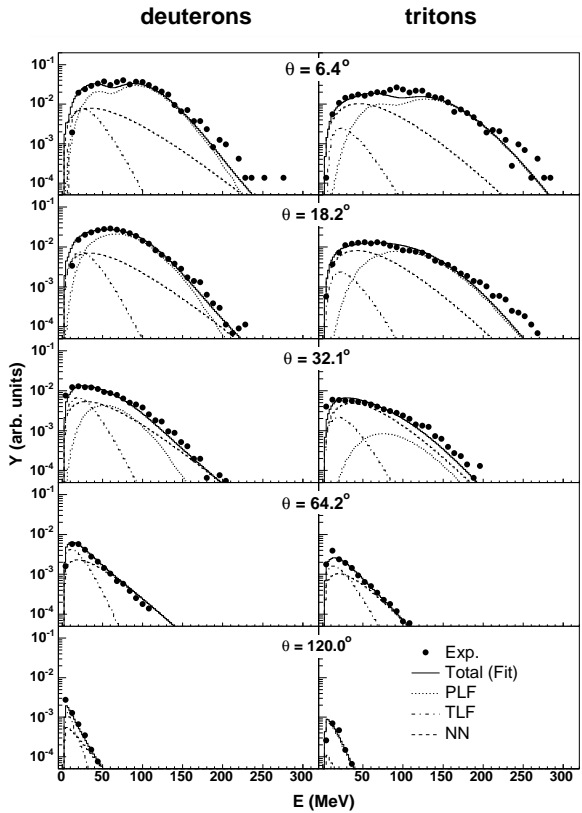}
\caption{\footnotesize  The three source fits for deutrons (left
panels) and tritons (right panels) in ($Bin2$) events  of
$^{40}$Ar + $^{58}$Ni reaction. The lab angle is $6.4^\circ$,
$18.2^\circ$, $32.1^\circ$, $64.2^\circ$ and $120^\circ$,
respectively, from top to bottom. The meanings of the symbols and
lines are depicted in the bottom right panel. } \label{fig_3sfit}
\end{figure}

Fig.~\ref{fig_3sfit} shows examples of the three source fits for
deuterons and tritons in the second most violent  bin ($Bin2$).
From these fits we know the relative contributions from the of QP,
NN and QT sources.  Employing this information to determine the
energy and angular dependent probabilities we analyze the
experimental events once again and, on an event by event basis,
use a Monte Carlo sampling method to assign each LCP  to one of
the sources QP, or NN, or QT. For example, the probability that a
certain  LCP $i$ ({\it i.e.} $p$, $d$ and  $t$ etc.)  will be
assigned to the QP source is
\begin{equation}
Prob^{QP}(E_{lab},\theta,i) = \frac{(\frac{d^2N}{dE_{lab}
d\Omega_{lab}})_L^{QP} }{ (\frac{d^2N}{dE_{lab} d\Omega_{lab}})_L}.
\end{equation}

\begin{figure}
\vspace{-.2truein}
\includegraphics[scale=0.45]{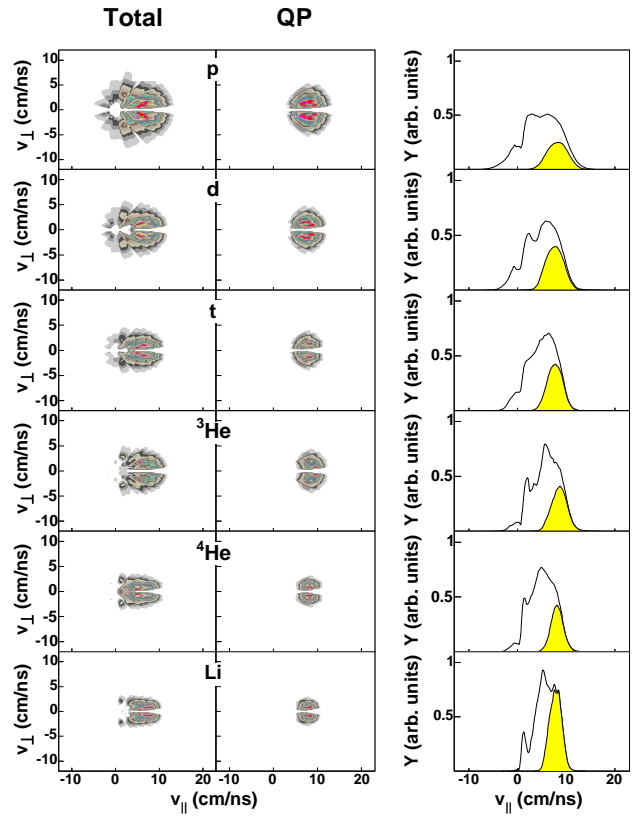}
\caption{\footnotesize  (Color online) The velocity contour plots
for the light charged particles  in violent events ($Bin2$) of
$^{40}$Ar + $^{58}$Ni reaction. From top to bottom,  protons,
deutrons, tritons, 3He, $\alpha$ and Lithium ; the left panel
shows the total velocity contour plot,  the middle column depicts
the velocity contour plots of particles from the  QP source and
the right column presents the corresponding distributions of the
parallel velocity from the total contribution (solid histogram)
and from the  QP source (hatched area). See details in text. }
\vspace{1.0truein} \label{fig_v1v2}
\end{figure}

To illustrate the results of such a procedure, we show, in
Fig.~\ref{fig_v1v2}, the velocity contour plots for protons to
Lithium associated with the highest multiplicity windows in the
$^{40}$Ar + $^{58}$Ni reaction.

The panels on the left represent the data before any  selection,
{\it i.e.} the raw data with contributions from all emission
sources. Obviously the particles are of mixed origin  and it is
difficult  to make a meaningful physical analysis. Since we are
interested in the QP source, we show, in the middle panels, the
velocity contour plots for particles assigned to the QP source
using the above reconstruction method. As expected from the
technique employed, the results exhibit clean, nearly spherical,
velocity contours,  corresponding to isotropic emission in the
rest frame of QP source.

The projected parallel velocity distributions are depicted in the
right panels of  Fig.~\ref{fig_v1v2}. The solid histograms
represent the total distribution  and the histogram with the
hatched area represents the contribution from the QP source. The
peak velocity of this QP contribution is close to the initial
projectile velocity although some dissipation is evident.

Intermediate mass fragments, IMF,  with Z$\geq$4 were identified
in the telescope modules of NIMROD. For such ejectiles we have we
have not used such fitting techniques. Rather we have used a
rapidity cut ($>0.65$ beam rapidity) to assign IMF  to the QP
source. We also checked the sensitivity of the above rapidity cut
to the results, eg., using  $>0.55$ or  $>0.75$ beam rapidity,
there are only minor changes for source mass, excitation energy
and temperatures etc, within $\sim 10 \%$ error bars. This has no
influence on any conclusions we draw in this article. Of course,
this is expectable for such a binary-dominated reaction mechanism.
Once we have identified all LCPs and IMFs which are assumed to
come from the QP source, we can reconstruct the whole QP source on
an event-by-event basis.

\begin{figure}
\includegraphics[scale=0.3]{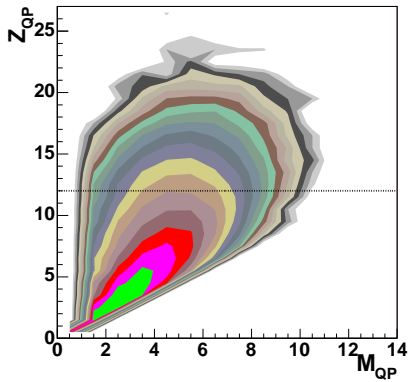}
\caption{\footnotesize  (Color online) The correlation of total
charge number of QP ($Z_{QP}$) and the total multiplicity of
charged particle ($M_{QP}$) for the  violent events of $^{40}$Ar +
$^{58}$Ni reaction. The region above the dotted line is used to
reconstructed the quasi-projectile. Note that the scale of z-axis
is logarithmic.} \label{fig_Zqp_Mqp}
\end{figure}

Fig.~\ref{fig_Zqp_Mqp} shows the two dimensional plot of total
charged particle multiplicity ($M_{QP}$)  and total atomic number
($Z_{QP}$) for the  reconstructed QP source. Due to the limited
geometrical coverage of the telescopes, the efficiency  of
detection for nearly complete  QP events is low. Note that the
scale of z-axis of the figure is logarithmic.

To select the nearly complete QP events, we  choose events with
$Z_{QP} \geq 12$ ({\it i.e.} as good events. The part of the
distribution  above the line in Fig.~\ref{fig_Zqp_Mqp})
corresponds to the accepted region of violent collisions for the
$^{40}$Ar + $^{58}$Ni reaction. The reconstructed good events in
that region account for $4.3\%$ of the total events for the
violence bins selected ($Bin1$ and $Bin2$). For the reactions
$^{40}$Ar + $^{48}$Ti and  $^{40}$Ar + $^{27}$Al, a similar
portion of the good central events has been collected to make the
same analysis. Totally,  28000, 54000 and 56000 good QP events
have been accumulated to make the following analysis for $^{40}$Ar
+ $^{58}$Ni, $^{48}Ti$ and $^{27}Al$ reactions, respectively. For
this analysis the velocity of the QP source was determined, event
by event, from the momenta of all QP particles.

\section{General Properties of the Excited  QP}

After the reconstruction of the QP particle source,
the excitation energy was deduced event-by-event using the energy
balance equation \cite{Cussol},  where the kinetic energy of charged
particles (CP), the mass excesses and the (undetected) neutron contributions  were
considered. i.e.,
\begin{equation}
 E^* = \sum_{i=1}^{M_{tot}}E_i^{kin}(CP) + \frac{3}{2} M_n T + Q
\end{equation}
where $Q$, the mass excess of the QP system  is determined from
the mass difference between the  final QP mass, $A_{QP}$ and the
sum over the masses of the detected particles of the reconstructed
QP, $\sum_{i=1}^{M_{tot}} A_i(CP)$. $A_i(QP)$ is determined from
the total reconstructed charge of the QP, assuming the QP has the
same $N/Z$ as the  initial projectile and
 $A_i(CP)$ is the mass of each detected charged particle,  which
was calculated from the measured $Z_i(CP)$ through the numerical
inversion of EPAX parameterization \cite{EPAX}, except for Z = 1
and 2 for which experimental mass identification was achieved. The
neutron multiplicity $M_n$ was obtained as the difference between
the mass number ($A_{QP}$) of the QP and the sum of nucleons bound
in the detected charged particle, i.e., $M_n = A_{QP} - \sum
A_i(CP)$. $E_i^{kin}(CP)$ is the kinetic energy of the charged
particles in the rest frame of QP. The contribution of the neutron
kinetic energy was taken as 3/2$M_n$T with an assumed  T that is
equal to that of the protons. As our detector is not 100$\%$
efficient we corrected observed events (on the average) for
undetected mass and energy. For a particular excitation energy bin
the missing multiplicity for a given ejectile  is the difference
between the multiplicity derived from the source fit
 and the average detected ejectile multiplicity for events in the
 acceptance window. Assuming that missed particles have the same
 average kinematic properties as the detected particles of the same
 species allows the appropriate corrections to be made. Since almost
 complete projectile-like species were selected initially the
missing particles were usually protons. Using these techniques we
find that the average QP has a mass of 36 and a charge of 16.

Assuming the mean mass of missing particles in a  given $E^*/A$
window is equal to $\Delta A$ , the contribution of missing
excitation energy $\Delta E^*$ can be approximated as $\Delta E^*
= \Delta A \cdot \frac{E^*_{meas.}}{\sum A_{LCP}}$, where
$E^*_{meas.}$ is the excitation energy before the correction and
$\sum A_{LCP}$ is the sum of the masses of LCP ( A$\leq 7 $) and
neutrons in the same  $E^*/A$ window. Thus the real excitation
energy should be $E^*_{meas.} + \Delta E^* $. Filtering results of
AMD-GEMINI calculations \cite{Ma_prepare} by applying experimental
acceptances  leads to very similar corrections to those employed.

\begin{figure}
\vspace{-0.5truein}
\includegraphics[scale=0.3]{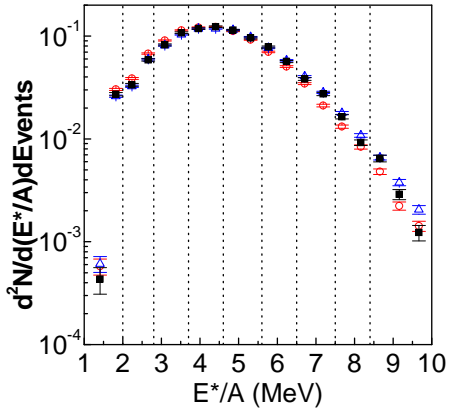}
\vspace{-0.5truein} \caption{\footnotesize  (Color online)
Excitation energy distribution of QP formed in $^{40}Ar$ +
$^{27}Al$ (open circles), $^{48}Ti$ (open triangles) and $^{58}Ni$
(solid squares). Dotted lines indicate the selected excitation
energy bins.} \label{fig_exc}
\end{figure}

Fig.~\ref{fig_exc} depicts  normalized excitation energy
distributions for the selected QP events in $^{40}Ar$ + $^{27}Al$
(open circles), $^{48}Ti$ (open triangles) and $^{58}Ni$ (solid
squares) for  $Bin1 + Bin2$. These distributions are very similar.
For  violent collisions, the highest excitation energy of the QP
can reach  9 MeV/nucleon. In the following analysis, we will
generally separate the excitation energy distributions into 9
windows, as shown by the slices in Fig.~\ref{fig_exc}. For
simplicity, we sometimes identify these $E^*/A$ windows as $Exc1$
through $Exc9$.

\begin{figure}
\vspace{-0.3truein}
\includegraphics[scale=0.45]{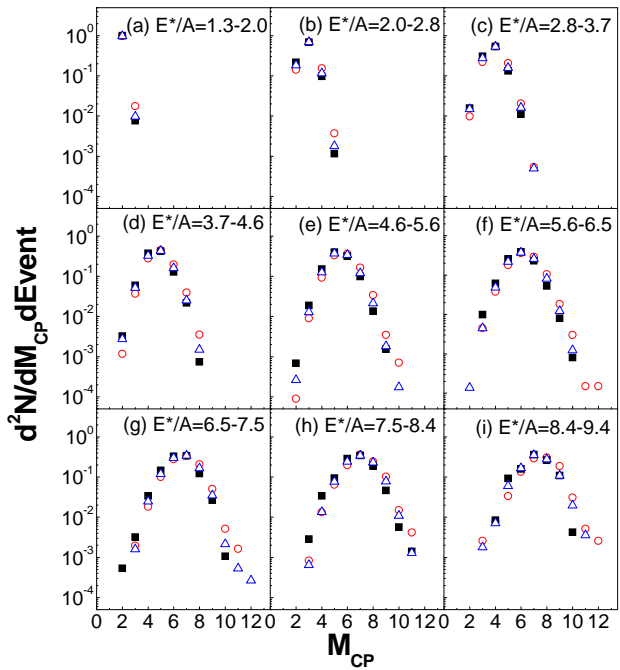}
\vspace{-0.8truein} \caption{\footnotesize  (Color online) The
total multiplicity distribution of charged particles from the QP
systems formed in $^{40}Ar$ + $^{27}Al$ (open circles), $^{40}Ti$
(open triangles) and $^{58}Ni$ (solid squares).  }
\label{fig_MtotQP}
\end{figure}

\begin{figure}
\includegraphics[scale=0.4]{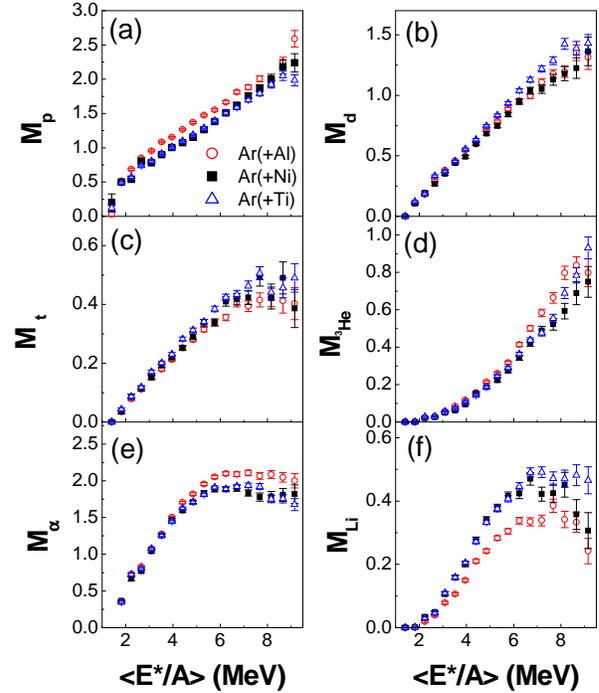}
\vspace{-0.3truein}
\caption{\footnotesize  (Color online) Average
multiplicity of $p$ (a), $d$ (b), $t$ (c), $^3He$ (d), $\alpha$
(e) and $Li$ (f) from the QP systems formed in $^{40}Ar$ +
$^{27}Al$ (open circles), $^{40}Ti$ (open triangles) and $^{58}Ni$
(solid squares) as a function of $E^*/A$.} \label{fig_mult}
\end{figure}

Fig.~\ref{fig_MtotQP} shows the total multiplicity distribution of
charged particles in 9 excitation energy windows. For the
quasi-projectiles formed in Ar induced reaction with different
targets, the distributions keep the nearly same which is a
reasonable results thanks of a clean QP reconstruction technique.
Fig.~\ref{fig_mult} shows average multiplicity of LCP as a
function of excitation energy. For p, d, t and $^3He$, the
multiplicity rises monotonically but for $\alpha$ and $Li$, the
multiplicities peak at $E^*/A$ near 6 MeV/nucleon  and then drop
at higher excitation energy. This behavior is similar to the rise
and fall behavior of IMF  yield observed in many previous
multifragmentation studies \cite{Ogilvie,Tsang93,Ma_PRC95}. Due to
the small size of our light system, this appears to occur for much
smaller fragments, and even to be reflected in the $A$ = 4 yields.

The QP formed in Ar + Al, Ar + Ti and Ar + Ni collisions are
almost identical indicating that we have a clean technique for
identifying the QP.

\section{Experimental Evidence  of  Critical Behavior}

We have used several techniques to look for evidence of possible
critical behavior in the A $\sim$ 36 system. These include a
Fisher droplet model analysis of the charge distributions,
searches for the region of maximal   fluctuations using  many
different observables  and tests of the fragment topological
structure.

\subsection{Fisher Droplet Model Analysis of Charge Distributions}

The Fisher droplet model has been extensively applied to the
analysis of multifragmentation since  the pioneering experiments
on high energy proton-nucleus collisions by the Purdue group
\cite{Purdue1,Purdue2,Minich82}. Relative yields of fragments with
$3 \leq Z \leq 14$ could be well described by a power law
dependence $A^{-\tau}$ suggesting  the disassembly of a system
whose excitation energy was comparable to its total binding energy
\cite{Purdue2}. The  extracted value of the power law exponent was
$2 \leq \tau \leq 3$, which is in a reasonable  range for critical
behavior \cite{Fisher}. The success of this approach suggested
that the multi-fragmentation of nuclei might be analogous to a
continuous liquid to gas phase transition observed in more common
fluids.

In the Fisher Droplet Model  the  fragment mass yield
distribution, Y(A) , may be represented as
\begin{equation}
Y(A) = Y_0 A^{-\tau} X^{A^{2/3}}Y^A,
\end{equation}
where $Y_0$, $\tau$, $X$ and $Y$ are parameters. However, at the
critical point, $X$ = 1 and $Y$ = 1 and the cluster distribution is
given by a pure power law
\begin{equation}
Y(A) = Y_0 A^{-\tau},
\end{equation}
The model predicts a critical exponent $\tau \sim$ 2.21.

\begin{figure}
\includegraphics[scale=0.45]{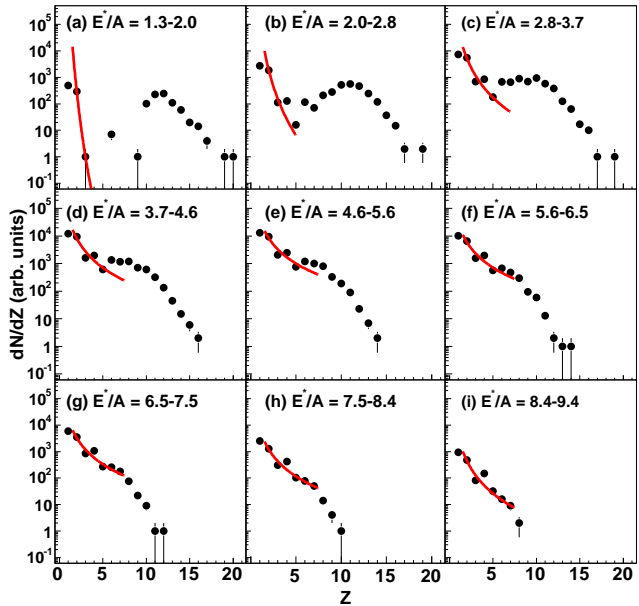}
\caption{\footnotesize  (Color online) Charge distribution of QP
in different $E^*/A$ window for the reaction  $^{40}$Ar +
$^{58}$Ni. Lines represent fits. } \label{fig_Zdist}
\end{figure}

In Fig.~\ref{fig_Zdist} we present,  for the QP from the reactions
of $^{40}Ar$ + $^{58}Ni$,  yield distributions,  $dN/dZ$, observed
for our nine intervals of excitation energy.

\begin{figure}
\vspace{-0.3truein}
\includegraphics[scale=0.35]{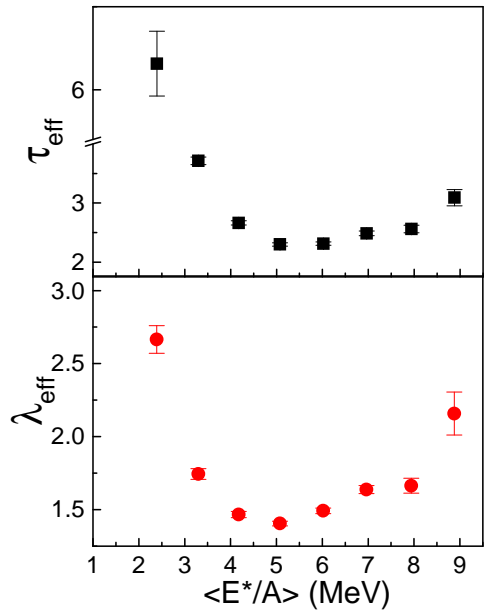}
\caption{\footnotesize  (Color online) $\tau_{eff}$ and
$\lambda_{eff}$ as a function of excitation energy for the QP
formed in $^{40}$Ar + $^{58}$Ni. } \label{fig_tau}
\end{figure}

At low  excitation energy a large $Z$ residue always remains,
$i.e.$ the nucleus is basically in the liquid phase accompanied by
some evaporated light particles. When $E^*/A$ reaches  $\sim$ 6.0
MeV/nucleon, this residue is much less prominent. As $E^*/A$
continues to increase, the charge distributions become steeper,
which indicates that the system tends to vaporize. To
quantitatively pin down the possible phase transition point, we
use a power law fit to the QP charge distribution in the range of
$Z$ = 2 - 7  to extract the effective Fisher-law parameter
$\tau_{eff}$ by
\begin{equation}
dN/dZ \sim Z^{-\tau_{eff}}.
\label{equ_tau_eff}
\end{equation}
The upper panel of Fig.~\ref{fig_tau} shows $\tau_{eff}$ vs
excitation energy, a minimum with $\tau_{eff}$ $\sim$ 2.3 is seen
to occur in the $E^*/A$ range of  5 to 6 MeV/nucleon. $\tau_{eff}$
$\sim$ 2.3, is close to the  critical exponent of the liquid gas
phase transition universality class as predicted by Fisher's
Droplet model \cite{Fisher}. The observed minimum  is rather
broad.

\begin{figure}
\includegraphics[scale=0.45]{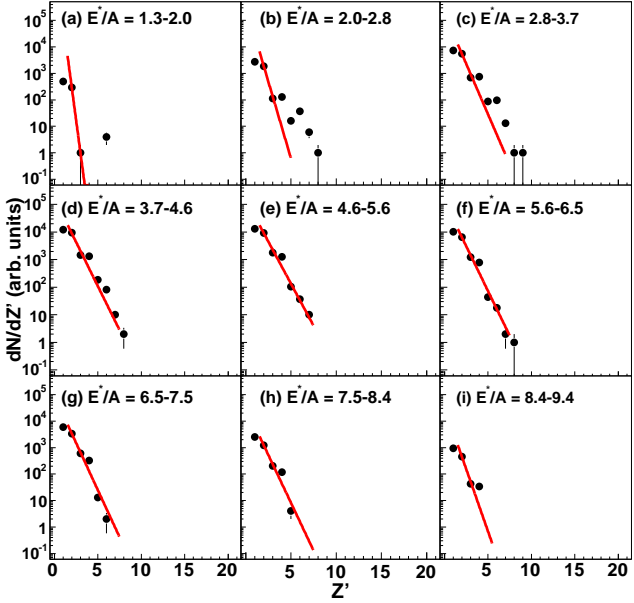}
\caption{\footnotesize  (Color online) Same as
Fig.~\ref{fig_Zdist} but the heaviest cluster is excluded on the
event by event basis. } \label{fig_noZmax}
\end{figure}

In a lattice gas model investigation of scaling and apparent
critical behavior, Gulminelli {\it et\ al.} have pointed out that,
in finite systems, the distribution of the maximum size cluster,
{\it i.e.} the liquid, might overlap with the gas cluster
distribution in such a manner as to mimic the critical power law
behavior with $\tau_{eff}$ $\sim$ 2.2 \cite{Gulminelli_2}. They
further note, however, that at that point the scaling laws are
satisfied, which suggests a potentially more fundamental reason
for the observation of the power law distribution
\cite{Gulminelli_2}. Assuming that the heaviest cluster in each
event represents the liquid phase, we have attempted to isolate
the gas phase by event-by-event removal  of the heaviest cluster
from the charge distributions. We find that the resultant
distributions are better described as exponential as seen in
Fig.~\ref{fig_noZmax}.

The fitting parameter $\lambda_{eff}$ of this exponential form $
exp(-\lambda_{eff} Z')$  was  derived and is  plotted against
excitation energy in the lower panel of Fig.~\ref{fig_tau}. A
minimum is seen in the same region where $\tau_{eff}$ shows a
minimum.

\subsection{Maximal fluctuations}

\subsubsection{ Campi Plots}

One of the well known characteristics of the systems undergoing a
continuous phase transition is the occurrence  of the largest
fluctuations. These large fluctuations in cluster size and
density of the system arise because of the disappearance of the
latent heat at the critical point. In macroscopic systems such
behavior gives rise to the phenomenon of critical opalescence
\cite{Stanley}.

Campi suggested the use of event by event scatter  plots of the
natural log of the size of the largest cluster, ln$A_{max}$ versus
the natural log of the normalized second moment, ln$S_2$, of the
cluster distribution with the heaviest fragment removed. For our
analysis we use $Z{max}$ as the measure of the size of the largest
cluster and

\begin{equation}
S_2 = \frac{ \sum_{ Z_i \neq Z_{max}} {Z_i}^2 \cdot n_i (Z_i) }
{\sum_{ Z_i \neq Z_{max}} {Z_i} \cdot n_i (Z_i)}.
\end{equation}
where $Z_i$ is the charge number of QP clusters and $n_i(Z_i)$ is
the multiplicity of the cluster $Z_i$. Campi plots  have proved
to be very instructive in previous searches for critical
behavior \cite{Campi}.

\begin{figure}
\includegraphics[scale=0.45]{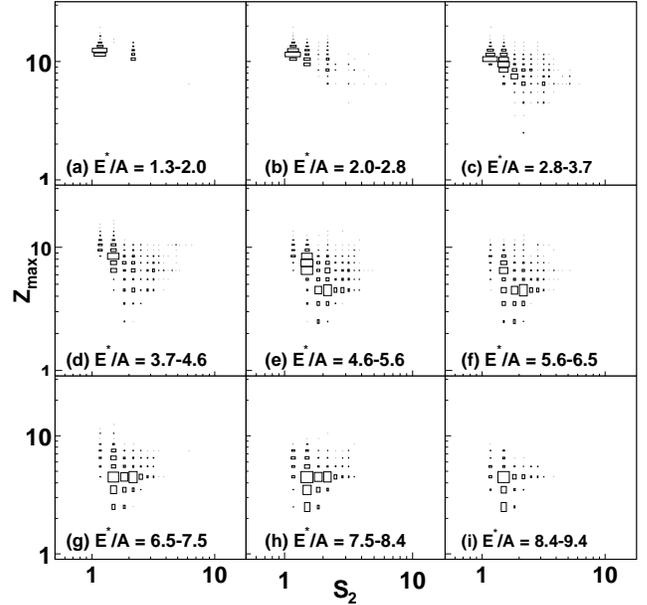}
\caption{\footnotesize  The Campi  plot for different excitation
energy windows  for the QP formed in $^{40}$Ar + $^{58}$Ni. }
\label{fig_campi}
\end{figure}

In Fig.~\ref{fig_campi} we present such plots for the nine
selected excitation energy bins. In the low excitation energy bins
of $E^*/A$ $\leq$ 3.7 MeV/nucleon,   the upper (liquid phase)
branch is strongly dominant   while at $E^*/A$ $\geq$ 7.5
MeV/nucleon,    the lower $Z_{max}$ (gas phase) branch is strongly
dominant. In the region of intermediate  $E^*/A$   of 4.6- 6.5
MeV/nucleon, the transition from the liquid dominated branch to
the vapor branch occurs, indicating that the region of maximal
fluctuations is to be found in that range.

Using the general definition of the $k$th moment as
\begin{equation}
M_k = \sum_{Z_i \neq Z_{max}} {Z_i}^k \cdot n_i (Z_i) .
\label{eq_Mk}
\end{equation}
Campi also suggested that the quantity, $\gamma_2$, defined as
\begin{equation}
\gamma_2 = \frac{M_2 M_0}{M_1^2},
\end{equation}
where $M_0$, $M_1$ and $M_2$ are the zeroth moment,  first moment
and second moment of the charge distribution, could be employed to
search for the critical region. In such an analysis, the position
of the maximum $\gamma_2$ value is expected to define the critical
point, {\it i.e.}, the critical excitation energy $E^*_c$, at
which the fluctuations in fragment sizes are the largest.

\begin{figure}
\vspace{-0.4truein}
\includegraphics[scale=0.30]{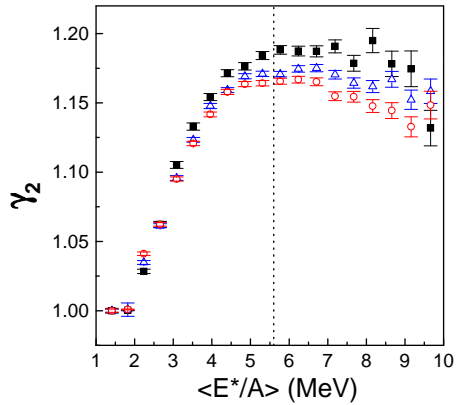}
\vspace{-0.4truein} \caption{\footnotesize  (Color online)
$\gamma_2$ of the QP systems formed in Ar + Al (open circles), Ti
(open triangles) and Ni (solid squares) as a function of
excitation energy.} \label{fig_g2}
\end{figure}

The excitation energy dependence of the average values of
$\gamma_2$  obtained in an event-by-event analysis of our data are
shown in Fig.~\ref{fig_g2}. $\gamma_2$ reaches its maximum in the
5-6 MeV excitation energy range. In contrast to observations for
heavier systems \cite{Elliott_PRC03}, there is no well defined
peak in $\gamma_2$ for our very  light system and $\gamma_2$ is
relatively constant at higher excitation energies. We note also
that the peak $\gamma_2$ value is lower than 2 which is the
expected smallest value for critical behavior in large systems.
However, 3D percolation studies  indicate that finite size effects
can lead to a decrease of $\gamma_2$ with system size
\cite{Campi92a,Campi92b}. For a percolation system with 64 sites,
peaks in $\gamma_2$ under two are observed. Therefore, the lone
criterion $\gamma_2 > 2$ is not sufficient to discriminate whether
or not the critical point is reached. To carry out  further
quantitative explorations of maximal fluctuations we have
investigated several other proposed observables expected to be
related to fluctuations and to signal critical behavior. These are
discussed below.

\subsubsection{ Fluctuations in the distribution of $Z_{max}$ }

It is supposed that the cluster size distributions should manifest
the maximum fluctuations around the critical point where the
correlation length diverges.  As a result of constraints placed by
mass conservation, the size of the largest cluster should then
also show  large fluctuations \cite{Stanley}. Thus, it has been
suggested that a possible signal of critical behavior is the
fluctuation in the size of the maximum fragment \cite{Campi}.
Recently, Dorso et al. employed a molecular dynamics model to
investigate fluctuations in  the atomic number of the heaviest
fragment ($Z_{max}$)  by determining its normalized variance
($NVZ$) \cite{Dorso},
\begin{equation} NVZ = \frac{\sigma^2_{Z_{max}}}{\langle Z_{max}\rangle  }
\end{equation}

\begin{figure}
\vspace{-0.3truein}
\includegraphics[scale=0.45]{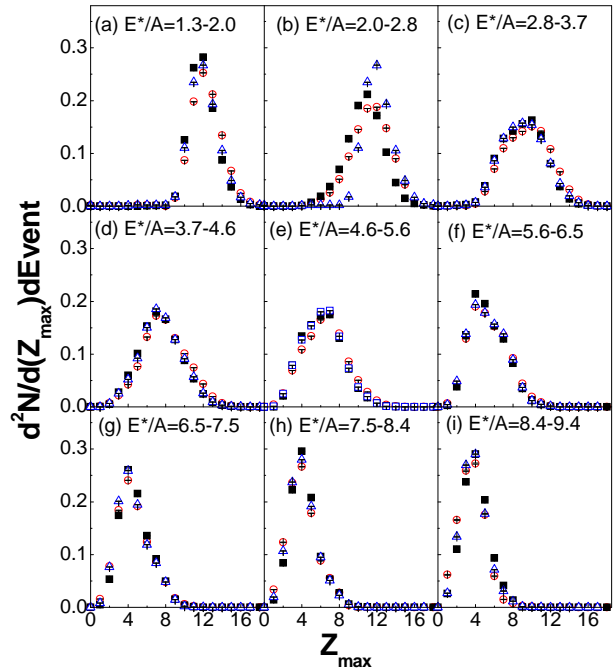}
\vspace{-0.5truein}\caption{\footnotesize  (Color online)
$Z_{max}$ distributions of the QP systems formed in Ar + Al (open
circles), Ti (open triangles) and Ni (solid squares) in different
excitation energy windows. } \vspace{-0.1truein} \label{fig_Zmax}
\end{figure}

In that work, they performed  calculations of the $NVZ$ on two
simple systems, one of which should not exhibit critical behavior
and one which does. For the first they used a random partition
model in which the population of the different mass numbers is
obtained by randomly choosing values of $A$ following a previously
prescribed mass distribution \cite{Elattari}. In this case the
fluctuations in the populations are of statistical origin or are
related to the fact that the total mass $A_{tot}$ is fixed. No
signal of criticality is to be expected. In the second case they
explored the disassembly of systems of the same size employing a
finite lattice bond percolation model. Such a case is known to
display  true critical behavior \cite{Dorso}. They found that that
$NVZ$ peaks close to the critical point in the percolation model
calculation  but shows no such peak in the random partition model
calculation. This indicates that  the mass conservation criterion,
by itself,  can not induce the peak of $NVZ$. The details can be
found in \cite{Dorso}.

\begin{figure}
\vspace{-0.2truein}
\includegraphics[scale=0.3]{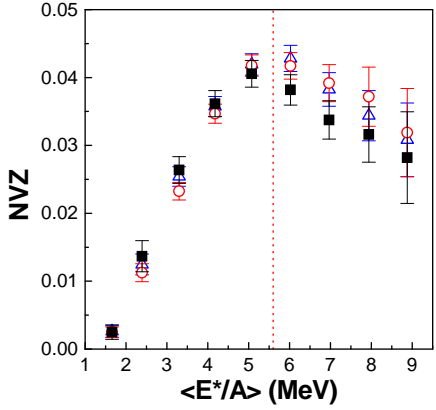}
\vspace{-0.3truein} \caption{\footnotesize  (Color online) $NVZ$
of the QP systems formed in Ar + Al (open circles), Ti (open
traingles) and Ni (solid squares) as a function of excitation
energy. Vertical line is at 5.6 MeV/u. See text.} \label{fig_NVZ}
\end{figure}

For our data we plot the normalized variance of $Z_{max}/Z_{QP}$
as a function of excitation energy in Fig.~\ref{fig_NVZ}. A clear
maximum, characterizing  the largest fluctuation of this order
parameter, is located in the  $E^*/A$ $\sim$ 5-6 MeV/nucleon,

\subsubsection{ Fluctuations in the  distribution of total kinetic  energy }

\begin{figure}
\vspace{-0.2truein}
\includegraphics[scale=0.3]{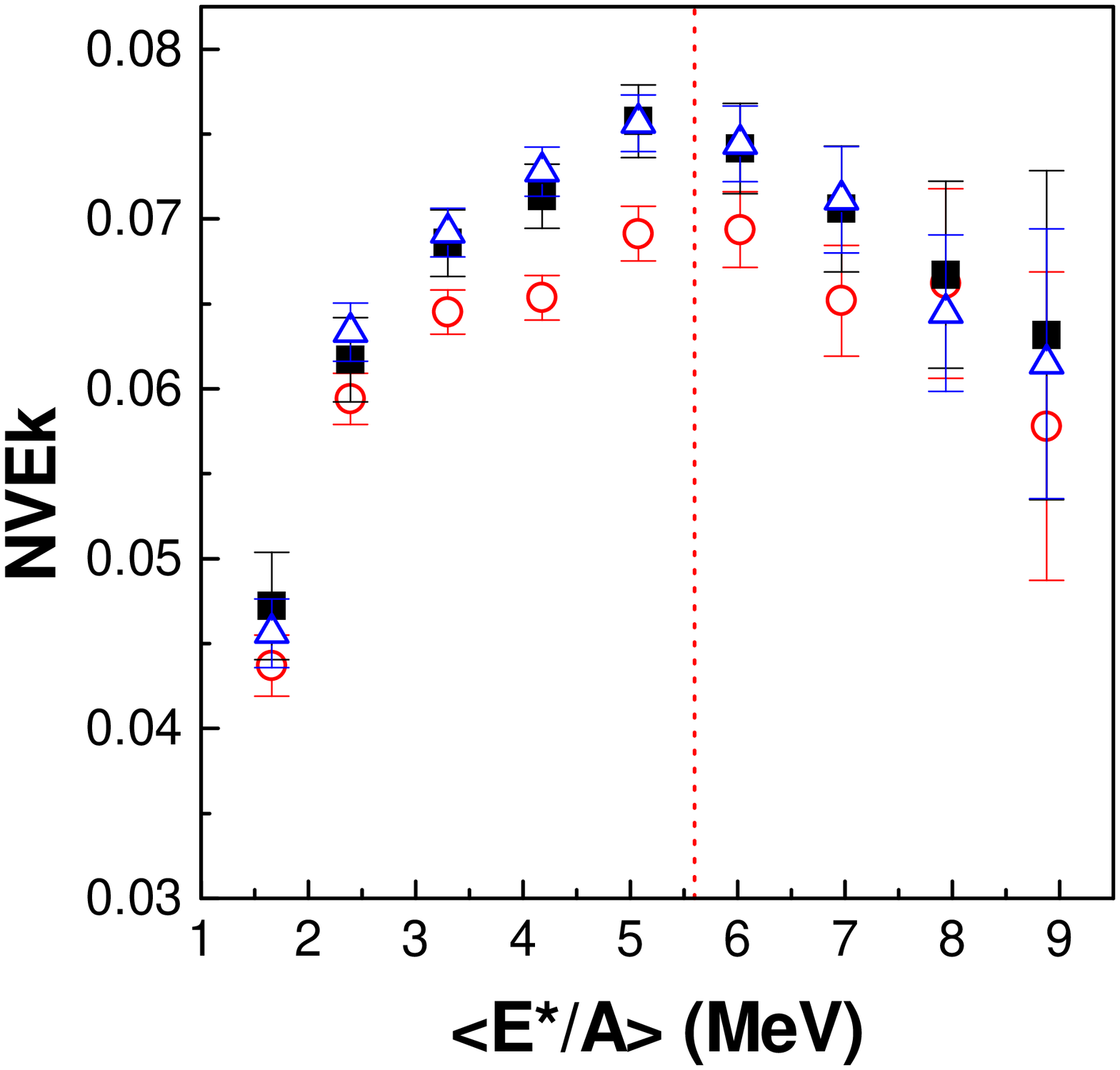}
\vspace{-0.3truein} \caption{\footnotesize  (Color online) $NVE$
of the QP systems formed in Ar + Al (open circles), Ti (open
traingles) and Ni (solid squares) as a function of excitation
energy. Vertical line is at 5.6 MeV/u. See text. }
\label{fig_NVEk}
\end{figure}

The system which we have studied  is a hot system.  If  critical
behavior occurs, it should also be reflected in large  thermal
fluctuations.  Using a definition similar to that of the
normalized variance of $Z_{max}$, we can define the normalized
variance of total kinetic energy per nucleon,
\begin{equation}
NVE = \frac{\sigma^2_{E^{tot}_{kin}/A}}{\langle E^{tot}_{kin}/A\rangle},
\end{equation}
where $E^{tot}_{kin}/A$ is the total kinetic energy per nucleon
and $\sigma_{E^{tot}_{kin}/A}$ is its width. Fig.~\ref{fig_NVEk}
shows the $NVE$ as a function of excitation energy. The observed
behavior is very similar  to that of $NVZ$. Again, the maximal
fluctuation was found at $E^*/A$ = 5 - 6 MeV/u. The maximal
thermal fluctuations are found in the same region as the maximal
fluctuations in cluster sizes.

The use of kinetic energy fluctuations  as a tool to measure
microcanonical heat capacities  has also been proposed
\cite{Chomaz_NPA99, Chomaz00,Agnostino}.

Based on the relation of the heat capacity per
nucleon and  kinetic energy  fluctuations, {\it i.e.}
\begin{equation}
\frac{c_V}{A_{QP}} = c \simeq c_K + c_I \simeq \frac{c_K^2}{c_K - A_{QP}
\sigma_k^2/T^2_m}
\label{eq_cV}
\end{equation}
where $c_K$ and $c_I$ are the kinetic and interaction
microcanonical heat capacities per particle calculated for the
most probable energy partition characterized by a microcanonical
temperature $T_m$ \cite{Chomaz_NPA99}. $T_m$ can be estimated by
inverting the kinetic equation of state \cite{Huang}
\begin{equation}
\langle E_{kin}^{tot}\rangle  = \langle \sum_{i=1}^M a_i\rangle
T_m^2 + \langle \frac{3}{2}(M - 1)\rangle  T_m .
\end{equation}
where $\langle \rangle$ indicates the average on the events with
the same $E_{kin}^{tot}$ and $a_i$ is the level energy density
parameter for fragment $i$ and $M$ is total multiplicity of QP
particles. A  negative heat capacity is indicated   if the kinetic
energy fluctuations exceed the canonical expectation
$A_{QP}\sigma_k^2/T_m^2 = c_K$.  In 35 MeV/nucleon Au + Au
collisions, the de- excitation properties of an Au QP formed at
excitation energies from 1 to 8 MeV/nucleon were investigated.
Abnormal kinetic energy fluctuations were observed near the
excitation energy  previously identified as the critical energy
\cite{Agnostino} and derived  negative heat capacities have been
taken  as a possible signal of the liquid gas phase transition.

\begin{figure}
\vspace{-0.5truein}
\includegraphics[scale=0.3]{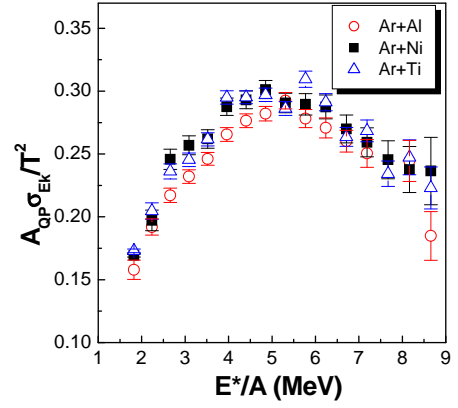}
\vspace{-0.5truein} \caption{\footnotesize  (Color online) The
kinetical energy fluctuation $A_{QP}\frac{\sigma^2_{kin}}{T^2_m}$
for the  QP formed in $^{40}Ar$ + $^{27}Al$ (open circles),
$^{48}Ti$ (open
triangles) and $^{58}Ni$ (solid squares). %Vertical line is at 5.6 MeV/u.
See text.}
\label{fig_cv}
\end{figure}

In Fig.~\ref{fig_cv} we present  the variable
$A_{QP}\frac{\sigma^2_{kin}}{T^2_m}$ as a function of excitation
energy as observed for the present system. The broad peak located
at $E^*/A$ = 4.0 - 6.5 MeV as Fig.~\ref{fig_NVEk} indicates the
region of the largest kinetic fluctuations. We note that  the
value of this quantity never reaches 3/2,  which is the canonical
expectation. It is possible that for such a very small system the
finite size effects will limit this parameter to values well below
the canonical expectation as it does for $\gamma_2$. Hence the
quantitative value of heat capacity $c_V$ will be difficult to
derive using   Eq.~\ref{eq_cV} without $\it {a\ priori}$ knowledge
of $c_K$.  It is clear that any value of $c_K$ below 0.29 would
lead to an  apparent negative heat capacity.

\subsubsection{ Universal Scaling Laws: $\Delta$-scaling}

The recently developed theory of universal scaling laws for
order-parameter fluctuations has been advanced as providing a
method to select order parameters and characterize critical and
non-critical behavior, without any assumption of equilibrium
\cite{Botet_PRE}. In this framework, universal $\Delta$-scaling
laws of the normalized probability distribution $P[m]$ of the
order parameter $m$ for different "system size" $\langle
m\rangle$, should be observed:
\begin{equation}
\langle m\rangle  ^\Delta P [m]
= \Phi(Z_{(\Delta)}) \equiv \Phi[\frac{m-m^*}{\langle m\rangle  ^\Delta}],
\end{equation}
with $0 <\Delta \leq 1$, where $\langle m\rangle  $ and $m^*$ are
the average and the most probable values of $m$, respectively, and
$\Phi(z_{(\Delta)})$ is the (positive) defined scaling function
which depends only on a single scaled variable $Z_{(\Delta)}$. If
the scaling framework holds, the scaling relation is valid
independent  of any phenomenological reasons for changing $\langle
m\rangle  $ \cite{Botet_PRE}. The $\Delta$-scaling analysis is
very robust and can be studied even in small systems if the
probability distributions $P[m]$ are known with a sufficient
precision.

Botet {\it et al.} applied this universal scaling method to INDRA
data for $^{136}$Xe + $^{124}$Sn collisions in the range of
bombarding energies between 25 MeV/nucleon and 50 MeV/nucleon. As
the relevant order parameter they chose the largest fragment
charge, $Z_{max}$. It was found that, at $E_{lab}$ $\geq$ 39
MeV/nucleon, there is a  transition in the fluctuation regime of
$Z_{max}$. This  transition is compatible with a transition from
the ordered phase ($\Delta$ = 1/2) to the disordered phase
($\Delta$ = 1) of excited nuclear matter \cite{Botet_PRL01}. From
this study, they attributed the fragment production scenario  to
the family of aggregation scenarios which includes both
equilibrium models, such as the Fisher droplet model, the Ising
model, or the percolation model and non-equilibrium models, such
as the Smoluchowski model of gels. For such scenarios the average
size of the largest cluster, $\langle Z_{max}\rangle$,  is the
order parameter and the cluster size distribution at the critical
point obeys a power law with $\tau > 2$.

\begin{figure}
\includegraphics[scale=0.4]{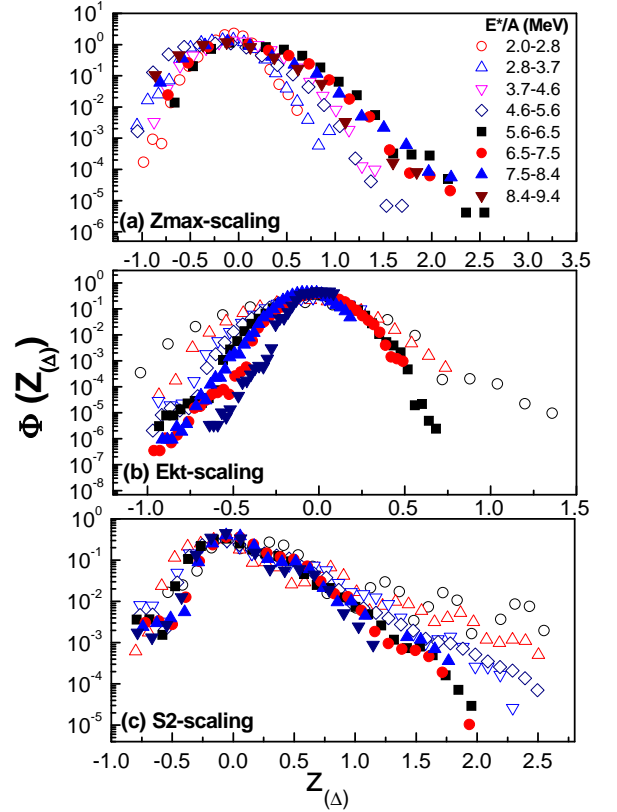}
\caption{\footnotesize (Color online) $\Delta$-scaling for
different quantities: the charge distribution of the largest
fragment (a), the total kinetic energy distribution per nucleon
$E^{tot}_{kin}/A$ (b) and the normalized second moment ($S_2$) in
different $E^*/A$ windows. The $\Delta$=1 scaling is generally
satisfied above 5.6 MeV/nucleon even it is not perfect in the
lower $Z_{(\Delta)}$ tail.} \label{fig_scaling}
\end{figure}

\begin{figure}
\vspace{-0.5truein}
 \includegraphics[scale=0.3]{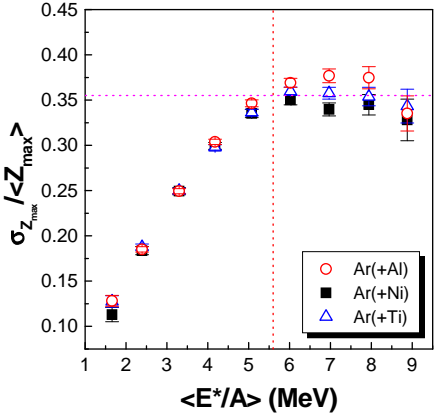}
 \vspace{-0.5truein} \caption{\footnotesize (Color online)
 $\sigma_{Z_{max}}/\langle Z_{max}\rangle$ as a function $\langle
 E^*/A\rangle$  for the QP formed in $^{40}$Ar + Al (open circles),
 Ni (filled squares) and Ti (open triangles).}
 \label{fig_sgm_zmax}
 \end{figure}

The upper panel in Fig.~\ref{fig_scaling} shows that
$\Delta$-scaling of $P[Z_{max}]$ distributions for all $E^*/A$
windows above 2.0 MeV with an assumed $\Delta$ = 1. For our light
system, our results show that the higher energy data are very well
scaled with $\Delta$ = 1 (even though not perfectly in the lower
$Z_{(\Delta)}$ tail) but the lower energy data are not. Similar
behaviors are also observed in other quantities, such as the total
kinetic energy per nucleon $E^{tot}_{kin}/A$
(Fig.~\ref{fig_scaling}(b)) and the normalized second moment $S_2$
(Fig.~\ref{fig_scaling}(c)) of QP. This indicates a transition to
$\Delta$ = 1 scaling in the region of $E^*/A$ = 5.6 MeV. This
corresponds to the fluctuations of the $Z_{max}$ growing with the
mean value, i.e. $\frac{\sigma_{Z_{max}}}{\langle Z_{max}\rangle}
\sim $ constant (see Fig.~\ref{fig_sgm_zmax}). The saturation of
the reduced fluctuations of $Z_{max}$ (i.e.
$\frac{\sigma_{Z_{max}}}{\langle Z_{max}\rangle}$) observed above
corresponds to the transition to the regime of maximal
fluctuations \cite{Frankland1}. However the lower energy data are
not well scaled by $\Delta$ = 1/2.

\begin{figure}
\vspace{-0.2truein}
\includegraphics[scale=0.35]{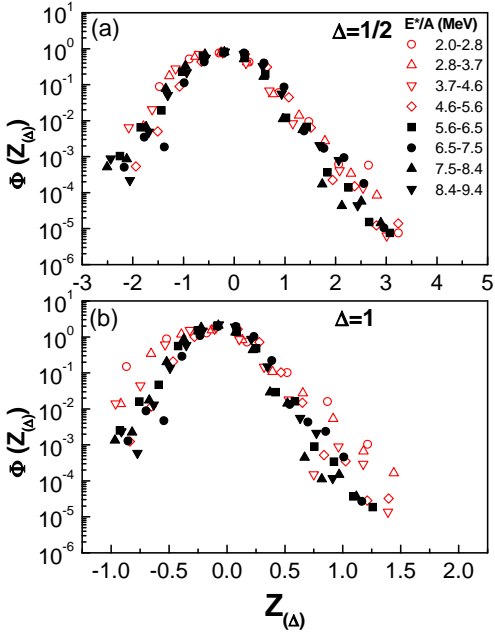}
\caption{\footnotesize (Color online) $\Delta$ = 1/2 scaling of
charged fragment multiplicity distributions in different $E^*/A$
windows for   the QP formed in $^{40}$Ar + $^{58}$Ni. }
\label{fig_mlt_scaling}
\end{figure}

The pattern of charged fragment multiplicity distributions $P[n]$
does not show any significant evolution with the excitation energy
(Fig.~\ref{fig_mlt_scaling}), and the data are perfectly
compressible in the scaling variables of the $\Delta$ = 1/2
scaling, i.e., the multiplicity fluctuations are small in the
whole excitation energy range. The scaling features of
experimental $Z_{max}$ ( Fig.~\ref{fig_scaling} (a)) and
multiplicity probability distributions (
Fig.~\ref{fig_mlt_scaling} ) are complementary and allow one to
affirm that the fragment production in Fermi energy domain follows
the aggregation scenario, such as the Fisher droplet model, and
two phases of excited nuclear matter with distinctly different
patterns of $Z_{max}$ fluctuation. It appears that $Z_{max}$ is a
very good order parameter to explore the phase change
\cite{Frankland_2004}.

From the studies of this section, we conclude that the largest
fluctuation phase ($\Delta$ = 1) is actually reached above 5.6
MeV/nucleon of excitation energy.

\subsection{Fragment Topological Structure}

\begin{figure}
\vspace{-0.3truein}
\includegraphics[scale=0.3]{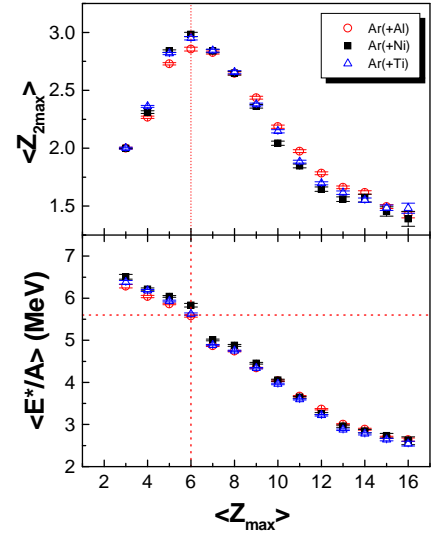}
\vspace{-0.2truein} \caption{\footnotesize (Color online) The
$\langle Z_{2max}\rangle$ as a function of $\langle
Z_{max}\rangle$. The mean excitation energy is shown beside of
points. } \label{fig_Z1Z2_proj}
\end{figure}

In addition to the thermodynamic and fluctuation features of the
system,  observables revealing some particular topological
structure may also reflect the critical behavior for a finite
system. For example, if we make a plot for the average value of
$Z_{2max}$ vs $Z_{max}$ in the different excitation energy
windows, we immediately see that a transition occurs near  5.6
MeV/nucleon (Fig.~\ref{fig_Z1Z2_proj}). Below that point $\langle
Z_{2max}\rangle$ increases with decreasing   $\langle
Z_{max}\rangle$. In these energy zones, the fragmentation is
basically dominated by evaporation  and sequential decay is
important. But above 5.6  MeV/nucleon excitation energy, $\langle
Z_{2max}\rangle$ decreases with decreasing   $\langle
Z_{max}\rangle$. In this region of excitation, the nucleus is
essentially fully vaporized and each cluster shows a similar
behavior.

Below we  present an exploration of  two more detailed observables
characterizing the topological structure, i.e.,  Zipf law
relationships and bimodality.

\subsubsection{ Zipf plots and Zipf's law }

Recently, Ma proposed  measurements of the fragment hierarchy
distribution as a means to search for the liquid gas phase
transition a finite system \cite{Ma_PRL99,Ma_EPJA}. The fragment
hierarchy distribution can be defined by the so-called Zipf plot,
i.e., a plot of the relationship between mean sizes of fragments
which are rank-ordered in size, i.e., largest, second largest,
etc. and their rank \cite{Ma_PRL99,Ma_EPJA}. Originally the Zipf
plot was used to analyze the hierarchy of usage of words in a
language \cite{Zipf}, {\it i.e.} the relative population of words
ranging from the word used most frequently to the word used least
frequently. The integer rank was defined starting from 1 for the
most probable  word and continuing to the least probable word.
Surprisingly, a linear relationship between the frequency and the
order of words was found. Later, many more applications of this
relationship were made in a broad variety of areas, such as
population distributions, sand-pile avalanches,  the size
distribution of  cities, the distribution in strengths of
earthquakes, the genetic sequence and the market distribution of
sizes of firms, etc. It has been suggested that the existence of
very similar linear hierarchy distributions in these very
different fields indicates that Zipf's law is a reflection of
self-organized criticality \cite{soc}.

\begin{figure}
\vspace{-0.7truein}
\includegraphics[scale=0.45]{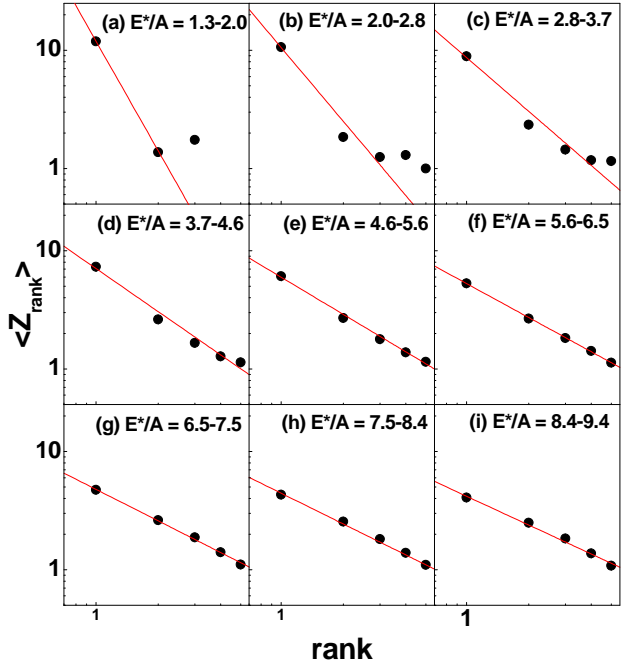}
\vspace{-0.8truein} \caption{\footnotesize (Color online) Zipf
plots in nine different excitation energy bins for   the QP formed
in $^{40}$Ar + $^{58}$Ni. The dots are data and the lines are
power-law fits (Eq.~\ref{eq_zipf}). The statistical error is
smaller than the size of the circles. } \label{fig_zipf}
\end{figure}

\begin{figure}
\includegraphics[scale=0.3]{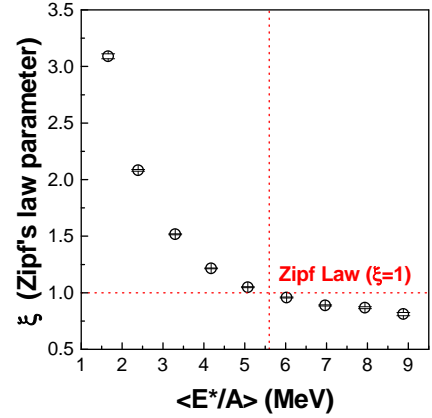}
\caption{\footnotesize (Color online) Zipf parameter as a function
of excitation energy  for   the QP formed in $^{40}$Ar +
$^{58}$Ni.  The position of cross illustrates the Zipf law is
reached around 5.6 MeV/u excitation energy. }
\label{fig_zipf_para}
\end{figure}

The significance of the 5-6 MeV region in our data is further
indicated by a Zipf's law analysis such as that proposed in
\cite{Ma_PRL99,Ma_EPJA}. In such an analysis, the cluster size is
employed as the variable to make a Zipf-type plot, and the
resultant distributions are fitted with a  power law ,
\begin{equation}
\langle Z_{rank}\rangle \propto  rank^{-\xi},
\label{eq_zipf}
\end{equation}
where $\xi$ is the Zipf's law parameter. In Fig.~\ref{fig_zipf} we
present Zipf plots for rank ordered average $Z$ in the nine
different energy bins. Lines in the figure are fits to the power
law expression of Eq.(\ref{eq_zipf}). Fig.~\ref{fig_zipf_para}
shows the fitted $\xi$ parameter as a function of excitation
energy. As shown in Fig.~\ref{fig_zipf}, this rank ordering of the
probability observation of fragments of a given atomic number,
from  largest to the smallest, does indeed lead to a Zipf's power
law parameter $\xi$ = 1 in  the 5-6 MeV/nucleon range.  When $\xi
\sim 1$, Zipf's law is satisfied. In this case, the mean size of
the second largest fragment is 1/2 of that of the the largest
fragment;  That of the third largest fragment is 1/3 of the
largest fragment, etc.

We note that the nuclear Zipf-type plot which was proposed in
Ref.~\cite{Ma_PRL99,Ma_EPJA} has been applied in the analysis of
CERN emulsion or Plastic data of Pb + Pb or Plastic at 158
GeV/nucleon and it was found that the nuclear Zipf law is
satisfied when the liquid gas phase transition occurs
\cite{Poland}.

\begin{figure}
\vspace{-0.2truein}
\includegraphics[scale=0.3]{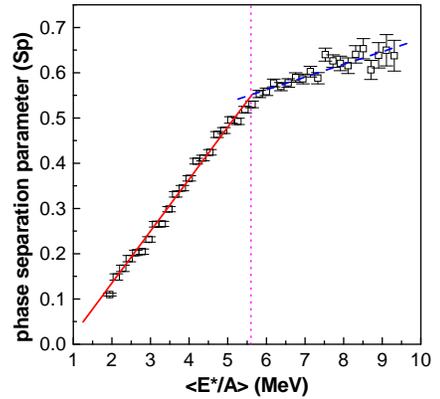}
\vspace{-0.2truein} \caption{\footnotesize (Color online) The
phase separation parameter as a function of excitation energy for
the QP formed in $^{40}$Ar + $^{58}$Ni. } \label{fig_Sp}
\end{figure}

In a related observation which is consistent with the formulation of Zipf's
law, percolation model calculations \cite{Cole} suggest that the
ratio $S_p$ = $\frac{\langle Z_{2max}\rangle}{\langle Z_{max}\rangle}$
reaches 0.5 around the phase separation point. Here   $Z_{2max}$ is the
atomic number of the second heaviest fragment in each event.
Fig.~\ref{fig_Sp} shows  $S_p$ versus $E^*/A$.
$S_p$ = 0.5 at 5.2 MeV/nucleon. It exhibits essentially
linear behavior (with two different slopes) above
and below that point.

\subsubsection{ Bimodality}

Another proposed test of phase separation is bimodality which was
suggested in \cite{Chomaz_PRE}. As has been noted \cite{Borderie}
this approach generalizes definitions based on curvature anomalies
of any thermodynamic potential as a function of an observable
which can then be seen as an order parameter. It interprets a
bimodality of the event distribution as  coexistence, each
component representing  a different phase. It provides a
definition of an order parameter as the best variable to separate
the two maxima of the distribution. In this framework when a
nuclear system is in the coexistence region, the probability
distribution of the order parameter is bimodal.

In analysis of INDRA data \cite{Borderie}, $(\frac{\sum_{Z_i \geq
13} Z_i - \sum_{3 \geq Z_i \leq 12} Z_i}{\sum_{Z_i \geq 3} Z_i})$
was  chosen as a sorting parameter. This parameter may be
connected with the density difference of the two phases
($\rho_l-\rho_g$), which is the order parameter for the liquid gas
phase transition.

\begin{figure}
\vspace{-0.3truein}
\includegraphics[scale=0.3]{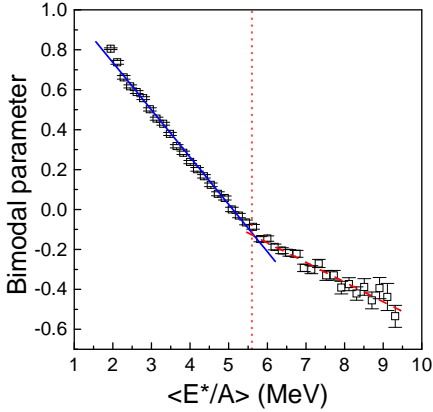}
\caption{\footnotesize (Color online) The average value of bimodal
as a function of excitation energy for   the QP formed in
$^{40}$Ar + $^{58}$Ni. } \label{fig_bimodal}
\end{figure}

For our very light system, if we consider the clusters
with $Z \leq 3$ as a gas and the clusters with $Z \geq 4$ as a liquid,
a parameter characterizing the bimodal nature of the distribution
can be defined  as
\begin{equation}
P = \frac{\sum_{Z_i \geq 4} Z_i - \sum_{Z_i \leq 3} Z_i}{\sum_{Z_i \geq 1} Z_i}.
\end{equation}
Fig.~\ref{fig_bimodal} shows the mean value of $P$ as a function of $E^*/A$.
Here again, the slope shows a distinct change at  $E^*/A$ = 5-6 MeV
where $P$ = 0, {\it i.e.} the point of equal distribution of $Z$
in the two phases.

\section{Caloric curve}

It is also interesting to ask how the caloric curve for this light
system behaves. Several different experimental methods have been
applied to the determination of  caloric curves for nuclear
systems. The most common of these are the use of slope parameters
of the kinetic energy spectra or the use of isotopic yield ratios
\cite{Albergo}. Since sequential decays  and side feeding  may be
important in either case, corrections  for such effects must
normally be applied to observed or "apparent" temperatures in
order to obtain the initial temperatures corresponding to the
initial excitation energies of the nuclei under investigation
\cite{ Majka_QSM, Durand}. For  heavier systems,  a number of
measurements  of caloric curves have been reported \cite{JBN0} and
references therein. In those measurements a flattening or
plateauing is generally  observed at higher excitation e4nergies.
For light systems such as the A $\sim$ 36 system studied here
there are relatively few  measurements of caloric curves.  For
construction of the caloric curve from the present data we have
used both the slope measurement and isotope ratio technique to
derive "initial temperatures" from the observed apparent
temperatures,  limiting the use of each to its own range of
applicability as discussed below.

\subsection{Low Excitation - the Liquid-Dominated Region}

\begin{figure}
\vspace{-0.3truein}
\includegraphics[scale=0.4]{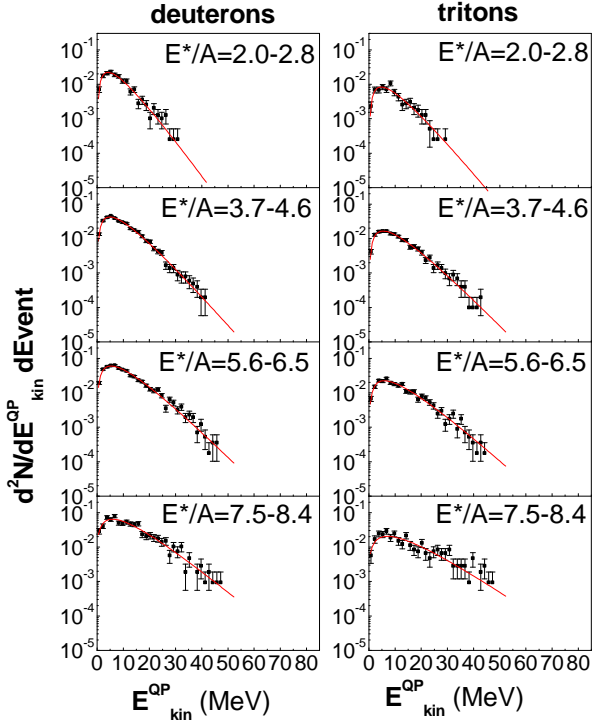}
\vspace{-0.3truein} \caption{\footnotesize (Color online) Kinetic
energy spectra in the QP frame of $^{40}$Ar + $^{58}$Ni in four
selected $E^*/A$ windows. Left panels are for deuteron and right
for tritons. The dots are experimental data and the lines are fits
with Eq.~\ref{eq_ekin}. } \label{fig_Eqp_dt}
\end{figure}

\begin{figure}
\vspace{-0.8truein}
\includegraphics[scale=0.4]{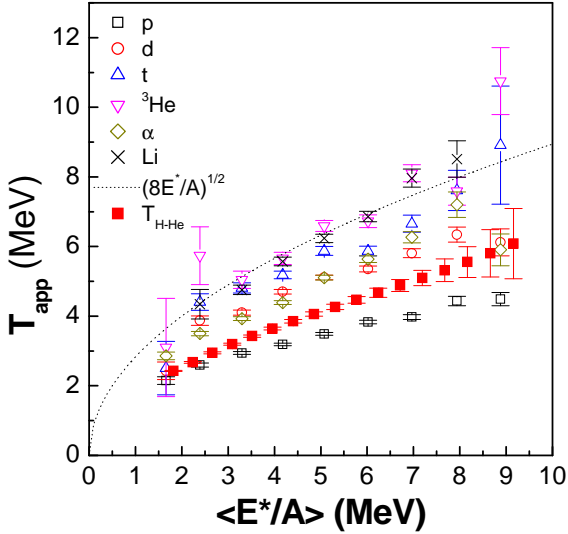}
\vspace{-0.5truein} \caption{\footnotesize (Color online) The
apparent temperatures from the slopes of different particles (open
symbols) and from isotopic ratio (solid squares) as a function of
excitation energy for   the QP formed in $^{40}$Ar + $^{58}$Ni.
The line is the Fermi gas model calculation: $T_s =
\sqrt{8E^*/A}$. } \label{fig_Ts_appar}
\end{figure}

Determinations based on spectral slope parameters  began with
fitting the kinetic energy spectra for different LCPs associated
with the nine different bins in excitation energy
 to obtain the apparent slope temperatures $T_s$ in the QP source frame.
$T_s$ can be obtained assuming a  surface emission type Maxwellian
distribution, {\it i.e.},
\begin{equation}
\frac{d^2N}{dE^{QP}_{kin}\cdot dEvent}
= c_0 \frac{E^{QP}_{kin} - V_{coul}} {T_{s}^2} exp(-\frac{E^{QP}_{kin}-
V_{coul}}{T_{s}}) .
\label{eq_ekin}
\end{equation}
where $E^{QP}_{kin}$ is the kinetic energy in the QP frame and
$V_{coul}$ is the barrier parameter. For an example,
Fig.~\ref{fig_Eqp_dt} shows the fits to the kinetic spectra of
deuterons and tritons in the QP frame in four different $E^*/A$
windows. The dashed lines represent the  fits and they show
excellent agreement with the data. Using such fitted results, the
excitation function of $T_s$ can be obtained for each light
charged particle as shown in  Fig.~\ref{fig_Ts_appar}. In this
figure, we also plot (dotted line) $T_s = \sqrt{8E^*/A}$ which
corresponds to the temperature from a simple Fermi gas assumption.
For these different particles, the apparent temperatures are
different from each other since the effects of sequential decay
are different for different particles. We note that the
temperatures, $T_s$, of $^3He$ and $Li$, are larger than those of
other LCPs, indicating that they might be the least affected by
the sequential decay effects, while $T_s$ for $protons$ shows
dramatically smaller values than the others indicating that the
$p$ spectra are strongly influenced by later stage emissions. We
then employed the measured excitation energy dependence of the
multiplicity for the ejectile under consideration to derive
initial temperatures from  the apparent slope  temperatures
\cite{hagel,Wada89}.

For  each LCP, the measured multiplicity is the sum over the
entire de-excitation cascade.  Since the temperature of an
evaporation residue in an excitation energy bin characterized by a
small change of  excitation from $E_1^*$ to  $E_2^*$ is, to a good
approximation,
\begin{equation}
\langle T_{ini}\rangle   = \frac{\langle M_2\rangle  \langle T_2\rangle
 - \langle M_1\rangle  \langle T_1\rangle  }{\langle M_2\rangle   - \langle M_1\rangle  }
 \label{eqn_KW}
\end{equation}
where $M_2$ and $M_1$ are the multiplicities of  a certain
LCP at the excitation energy  $E_2^*$ and   $E_1^*$ where
$E_1^* >  E_2^*$. The details of this method can be found
in references \cite{hagel,Wada89}.

\begin{figure}
\vspace{-0.7truein}
\includegraphics[scale=0.35]{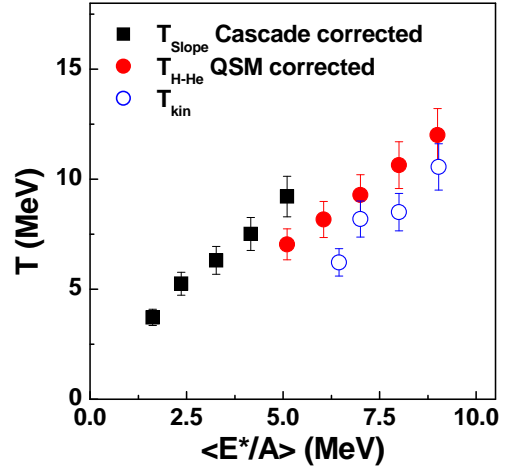}
\vspace{-0.4truein} \caption{\footnotesize (Color online) The
deduced caloric curves for   the QP formed in $^{40}$Ar +
$^{58}$Ni. The symbols are displayed in insert. See details in
text.} \label{fig_caloric}
\end{figure}

With this method, we can derive the initial temperatures for each
particle. For each particle except protons
 we obtained a reasonable
agreement of the respective initial temperatures and therefore use
their average values, as shown by solid squares in
Fig.~\ref{fig_caloric},  as a mean initial temperature for
plotting the caloric curve. For protons, the apparent temperature
is very low from fits as shown in Fig.~\ref{fig_Ts_appar}
 since  a large portion of protons may originate from the side
feeding besides the sequential decay chain. However,
  the former can not be corrected with Eq.~(\ref{eqn_KW}).
It must be emphasized that this technique is based on the
assumption of sequential evaporation of the ejectiles from a
cooling compound nucleus source \cite{hagel,Wada89}. Given that
the various observables discussed above suggest an important
transition at 5.6 MeV/u excitation energy, this method should not
be appropriate
 above that energy. In fact, initial temperatures deduced
using this approach exhibit a very rapid increase at excitation
energies above 6 MeV/nucleon (not shown). We take this as evidence that
sequential evaporation from a larger parent can not explain the
multiplicities in the higher energy region.  This is already
suggested by the energy dependence of various multiplicities in
Figure 8 as well as by much of the discussion in the previous
section.   Thus we do not employ this method based on slope
measurements above 5.6 MeV/nucleon .

\subsection {High Excitation (I) - the Vapor-Dominated Region }

The double isotope ratio temperature technique proposed by Albergo
{\it et\ al.}  \cite{Albergo} has been extensively discussed and
used in many experiments and theoretical calculations. Application
of this technique assumes that thermal equilibration and chemical
equilibration have been attained. In an experimental
determination, one of the major problems  is that secondary decay
effects can modify the initial temperature strongly
\cite{QSM,Durand,Majka_QSM}.

The experimental  apparent double isotope temperature, $T_{H-He}$,
can be deduced from the ratio of
$\frac{M_d/M_t}{M_{^3He}/M_\alpha}$:
\begin{equation}
T_{H-He} = \frac{14.3}{ln[1.59 (M_d \cdot M_\alpha) / (M_t \cdot M_{^3He})]} ,
\end{equation}
where $M_d$, $M_t$, $M_{^3He}$, and $M_\alpha$ are the isotopic
yields of $d$,$t$,$^3He$ and $\alpha$ from QP (see
Fig.~\ref{fig_mult}), respectively. As shown in
Fig.~\ref{fig_Ts_appar} (solid squares), the apparent isotopic
temperatures are well below those of the simple Fermi gas
assumption (dotted line in the figure) indicating a strong
influence due to secondary decay. To estimate the secondary decay
effect Quantum Statistical Model (QSM) calculations were performed
to correct the observed double isotope H-He ratio temperatures
($T_{H-He}$) for these effects.

For this purpose we compared results of two different
calculations, the first published in reference
\cite{Cussol,Durand} and the second carried out for this work
employing the QSM model described in reference
\cite{QSM,Durand,Majka_QSM}. The results of the two QSM models are
in quite good agreement with each other. For nuclei with $A \sim
36$ in the excitation energy range of Interest,  averaging results
of these models indicates that $T_{init} = (1.75 \pm 0.06) \times
T_{HHe}$. The corrected isotopic temperature is shown in
Fig.~\ref{fig_caloric} as the solid circles.

As emphasized above, this method is based on a model which assumes
simultaneous fragmentation of a reduced density equilibrated
nucleus and subsequent secondary evaporation from the primary
fragments \cite{QSM}. This method should be inappropriate in the
lower excitation energy where the vapor assumption of the QSM is
violated. In this case, we do not apply the technique below 5
MeV/nucleon

The  two techniques differ somewhat in the excitation energy range
near the transition point, indicating some systematic  error due
to using different techniques in the transition point region. This
supports the  argument for restricting the use of each  technique
to the "appropriate" excitation energy region.

We note that the caloric curve, defined in this manner, exhibits
no obvious plateau. A polynomial fit to the data points leads to a
temperature at the transition point of 8.3 $\pm$ 0.5 MeV.

\subsection {High Excitation (II) - the Ideal Vapor Assumption}

If the vapor phase may be characterized as an ideal gas of
clusters \cite{Fisher}, then, at and above $T$ = 8.3 MeV, this
should be signaled by a kinetic temperature, T$_{\rm{}kin} =
\frac{2}{3} {\rm E}^{\rm{}th}_{\rm{}kin}$, where
E$^{\rm{}th}_{\rm{}kin}$ is the Coulomb corrected average kinetic
energy of primary fragments. Secondary decay effects  make it
difficult to test this expectation. However, in an inspection of
the average kinetic energies or apparent slope parameters
(Fig.~\ref{fig_Ts_appar} ) for the different species observed , we
find that, {\it for each excitation energy window}, the average
kinetic energy  of $^{\rm{}3}$He isotropically emitted in the
projectile like frame, is higher than those of other species. This
together with simple model estimates indicates that the
$^{\rm{}3}$He spectra are the least affected by secondary decay.
Kinetic temperatures for $^{\rm{}3}$He, defined as
$\frac{2}{3}(\bar{E}_k-B_c)$ where $\bar{E}_k$ is the average
kinetic energy and $B_c$ is the Coulomb energy( obtained from the
fits), are plotted as open squares in Fig.~\ref{fig_caloric}.
Above $T$ = 8.3 MeV  the kinetic temperatures show a similar trend
to that of the the chemical temperatures but are approximately 1.5
MeV lower. While not perfect this approximate agreement provides
additional evidence for disassembly of an equilibrated system.

For heavier systems a plateau or flattening is often observed in
caloric curves \cite{JBN0} and the region of entry into the
plateau appears to be very close to the point which has been
identified as the point of maximal fluctuations. The reason for
this flattening is still under discussion. It may reflect
expansion and/or spinodal decomposition inside the coexistence
region \cite{Dorso_2,JBN3,Ono,Norenberg,Viola,Sob}. In contrast,
our light system does not show the flattening. This suggests which
that the transition under investigation may differ from that seen
in the heavier systems.

Taken together with the observations indicating maximal
fluctuations and the particular features of the fragment
topological  structure at 5.6 MeV/u excitation, the comportment of
this caloric curve provides further evidence suggesting that the
observed transition is taking place at, or very close to, the
critical point.

\section{Determination of Critical Exponents}

Since the pioneering work  on extraction of the critical exponents
for nuclear multifragmention from EOS data \cite{EOS_expo},
several additional experimental  and  theoretical efforts have
been attempted
\cite{Agostino_NPA99,Elliott,Elliott_PRC55,Elliott_PRC49}. In the
latter works, Elliott {\it et al.} show that the scaling behavior
can remain even in small systems and the critical  exponents can
be extracted.

\begin{figure}
%\vspace{-0.5truein}
\includegraphics[scale=0.4]{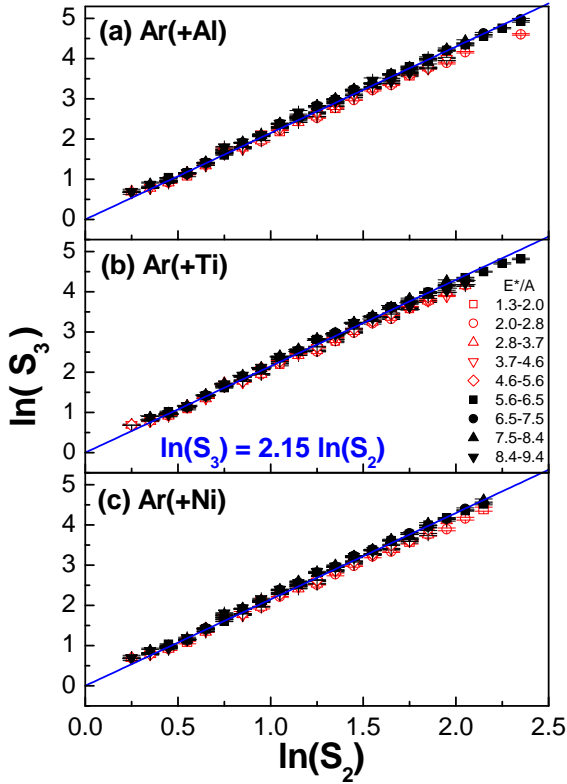}
\caption{\footnotesize (Color online) The correlation between
ln($S_3$) vs ln($S_2$) and the linear fit.} %\vspace{0.3truein}
\label{fig_tau_s23}
\end{figure}

In the Fisher droplet model, the critical exponent $\tau$ can be
deduced from the cluster distribution near the phase transition
point. In Sec.V(A), we already determined, from the yield
distributions,   $\tau_{eff} \sim 2.31 \pm 0.03$, which is close
to that for the liquid gas phase transition universality class. In
terms of the scaling theory, $\tau$ can also be deduced from,
($S_{corr}$), the slope of the correlation between ln($S_3$) vs
ln($S_2$) \cite{Agostino_NPA99}, where $S_3 = M_3/M_1$, shown in
Fig.~\ref{fig_tau_s23}, is related to $\tau$ as
\begin{equation}
\tau = \frac{3S_{corr}-4}{S_{corr}-1}.
\label{equ_tau_s23}
\end{equation}
Assuming the value of $T_c$ = 8.3 MeV as determined from our
caloric curve measurements, we explored the correlation of of
$S_2$ and $S_3$ in two ranges of excitation energy ¨C see Figure
31. The moments were computed by exclusion of  the species with
$Z_{max}$  in the "liquid" phase but inclusion in the "vapor"
phase.The slopes were determined from linear fits to the "vapor"
and "liquid" regions respectively and then averaged. In this way,
we obtained a   value of $\tau = 2.13 \pm 0.1$. See
Fig.~\ref{fig_tau_s23}.

Other  critical exponents can also be related to other moments of
cluster distribution, $M_k$, which were defined in
Eq.~\ref{eq_Mk}. Since, for our system, we have already deduced
the initial temperatures and determined a critical temperature
$T_c$ =  8.3 MeV  at point of maximal fluctuations, we can use
temperature as a control parameter for such determinations.  In
this context, the critical exponent $\beta$ can be extracted from
the relation
\begin{equation}
Z_{max} \propto (1-\frac{T}{T_c})^\beta,
\label{eq_beta}
\end{equation}
and the critical exponent $\gamma$ can be extracted from
the second moment via
\begin{equation}
M_2 \propto |1-\frac{T}{T_c}|^{-\gamma}.
\label{eq_gamma}
\end{equation}
In each, $|1-\frac{T}{T_c}|$ is the parameter which measures the distance
from the critical point.

\begin{figure}
\includegraphics[scale=0.4]{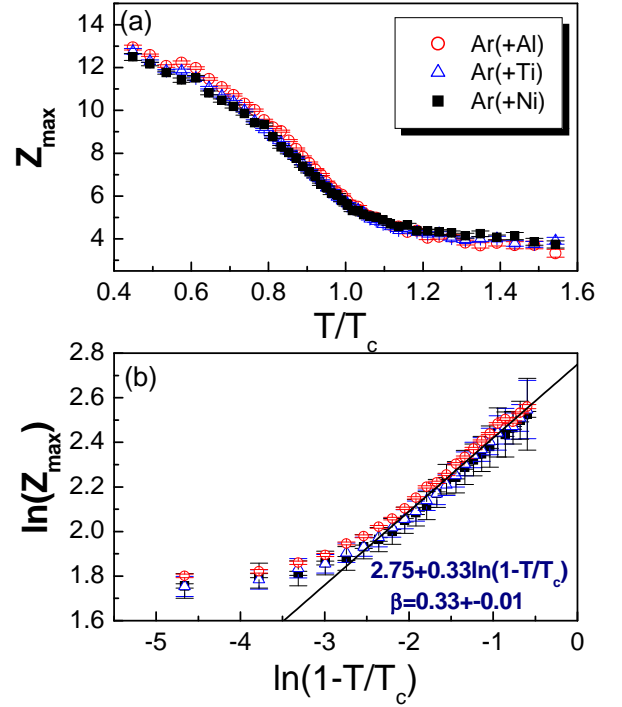}
\caption{\footnotesize (Color online) $Z_{max}$ as a function of
$T/T_c$ (a) and the extraction of the critical exponent $\beta$
(b). } \label{fig_beta}
\end{figure}

Fig.~\ref{fig_beta}  explores the dependence of $Z_{max}$ on $(
\frac{T}{T_c})$. We note a dramatic change of $Z_{max}$ around the
critical temperature  $T_c$.    LGM calculations also predict that
the slope of $Z_{max}$ vs $T$ will change at the liquid gas phase
transition \cite{Ma_JPG01}. Physically, the largest fragment is
simply related to the order parameter $\rho_l - \rho_g$ (the
difference of density in nuclear `liquid' and `gas' phases). In
infinite matter, the infinite cluster exists only on the `liquid'
side of the critical point. In finite matter, the largest cluster
is present on both sides of the phase transition point. In this
figure, the significant change of the slope of $Z_{max}$ with
temperature should correspond to a sudden disappearance of the
infinite cluster (`bulk liquid') near the phase transition
temperature. For the finite system, it  reflects the onset of
critical behavior there. Using the left side of  this curve ({\it
i.e. }liquid side), we can deduce the critical exponent $\beta$ by
the transformation of the $x$ axis variable to the distance from
the critical point. Fig.~\ref{fig_beta}b shows the extraction of
$\beta$ using Eq.~\ref{eq_beta}.  An excellent fit was obtained in
the region away from the critical point, which indicates a
critical exponent $\beta$ = 0.33 $\pm$ 0.01. Near  the critical
point, the finite size effects become stronger so that the scaling
law is violated. The extracted value of $\beta$ is that expected
for a liquid gas transition (See Table.II) \cite{Stanley}.

\begin{figure}
\vspace{-0.9in}
\includegraphics[scale=0.35]{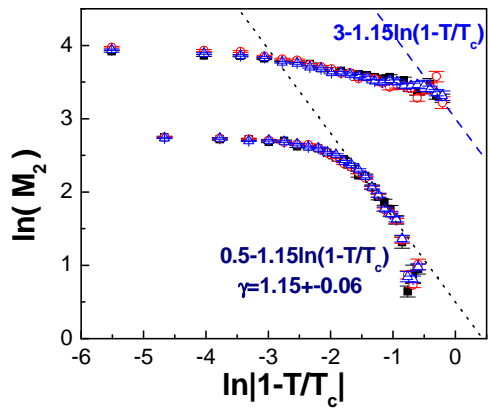}
\vspace{-0.6truein} \caption{\footnotesize (Color online) The
extraction of critical exponents $\gamma$. See texts for details.
} \label{fig_gamma}
%\vspace{-0.8truein}
\end{figure}

To extract the critical exponent $\gamma$, we take $M_2$ on the
liquid side without $Z_{max}$ but take $M_2$ on the vapor side
with $Z_{max}$ included. Fig.~\ref{fig_gamma} shows ln($M_2$) as a
function of ln($|1-\frac{T}{T_c}|$). The lower set of points is
from the liquid phase and the upper set of points is from the
vapor phase. For the liquid component, we center our fit to
Eq.~\ref{eq_gamma} about the center of the range of $(1-T/T_c)$
which leads to the linear fit and extraction of $\beta$ as
represented in Figure.~\ref{fig_beta}. We obtain the critical
exponent $\gamma$ = 1.15 $\pm$ 0.06. This value of $\gamma$ is
also close to the value expected for the
 liquid gas universality class (see Table II).
It is seen that the selected region has a good power law
dependence. However, a similar effort to extract the $\gamma$ in
the gas phase is not successful: a small value less than 0.20 is
deduced.  This may be due to the finite size effects for this very
light system. Since the largest cluster still exists in the vapor
side, its inclusion (or exclusion) in $M_2$ might perturb the
determination of  the moment, resulting in an imprecise value of
$\gamma$ extracted from the vapor phase. For comparison, we just
show, for the  vapor phase, a line representing the $\gamma$
derived from the liquid side. This line only agrees with the  last
few vapor points, {\it i.e. } the highest temperature points (the
contamination of $M_2$ should the least there).

Since we have the critical exponent $\beta$ and
$\gamma$, we can use the scaling relation
\begin{equation}
\sigma = \frac{1}{\beta + \gamma},
\end{equation}
to derive the critical exponent $\sigma$.
In such way, we get the $\sigma$ = 0.68 $\pm$ 0.04,
which is also very close to the expected critical exponent of
a liquid gas system.

Finally, it is possible to use the scaling relation
\begin{equation}
\tau = 2 + \frac{\beta}{\beta + \gamma},
\label{eq_tau}
\end{equation}
to check the $\tau$ value which was determined from the charge
distributions using Fisher droplet model power law fits around the
critical point (see Fig.~\ref{fig_Zdist}). Using Eq.~\ref{eq_tau}
we Obtain  $\tau$ = 2.22 $\pm$ 0.46, which, though less precise,
is in agreement with the values of $2.31\pm0.03$ obtained from the
charge distribution around the  point of maximal fluctuations and
$2.15 \pm 0.1$ extracted from the correlation of ln($S_3$) vs
ln($S_2$).

To summarize this section, we report in  Table.II  a  comparison
of our results with  the Values expected for the 3D percolation
and liquid gas system universality classes and with the results
obtained by Elliott et al. for a heavier system.  Obviously, our
values for this light system with A$\sim$36 are consistent with
the values of the liquid gas phase transition universality class
rather the 3D percolation class.

\begin{table}
\caption{Comparison of the Critical Exponents}
  \begin{tabular}{llll}
  \toprule
   Exponents & 3D Percolation & Liquid-Gas & This work\\
 \colrule
 &  &  & 2.22$\pm$0.46 (Eq.~\ref{eq_tau})\\
$\tau$ & 2.18 & 2.21 & 2.31$\pm$0.03 (Eq.~\ref{equ_tau_eff})\\
 &  &  & 2.13$\pm$0.10 (Eq.~\ref{equ_tau_s23})\\
$\beta$ & 0.41 & 0.33 & 0.33$\pm$0.01\\
$\gamma$ & 1.8 & 1.23 & 1.15$\pm$0.06\\
$\sigma$ & 0.45 & 0.64 & 0.68$\pm$0.04\\
   \colrule
 \botrule
\end{tabular}
\label{tab2}
\end{table}

\section{conclusions}

In conclusion, an extensive survey of the features of the
disassembly of  nuclei with  $A \sim 36$ has been reported. To
carry out this analysis, the de-excitation products of the A$\sim$
36 quasi projectile source were first reconstructed using a new
technique based upon on the three source fits to the light
particle spectra and use of a rapidity cut for IMF. Monte Carlo
sampling techniques were applied to assign all particles to one of
three sources (QP, NN and QT).

At an excitation energy $\sim$ 5.6 MeV/nucleon key observables
demonstrate the existence of maximal fluctuations in the
disassembly process. These fluctuation observables include the
Campi scattering plots and the normalized variances of the
distributions of order parameters, ($Z_{max}$) and total kinetic
energy.  Recently proposed $\Delta$-scaling analysis also show a
universal behavior at higher excitation energy where the
saturation of the reduced fluctuations of $Z_{max}$ ({\it i.e.}
$\frac{\sigma_{Z_{max}}}{\langle Z_{max}\rangle}$) is observed.
This  corresponds to the transition to a regime of large
fluctuations from an ordered phase at lower excitation energy.

At the same excitation energy $\sim$ 5.6 MeV/nucleon, the Fisher
droplet model prediction is satisfied, with a Fisher power law
parameter, $\tau$ = 2.3, close to the critical exponent of the
liquid gas phase transition universality class. In addition, the
fragment topological structure shows that the rank sorted
fragments obey Zipf's law,  proposed as a signature of liquid gas
phase transition \cite{Ma_PRL99}, at the maximal fluctuation
point. The related phase separation  parameter \cite{Cole} shows a
significant change of slope with excitation energy. The
correlation of the heaviest fragment and the second heaviest
fragment demonstrates a transition around 5.6 MeV/u of excitation
energy. A bimodality test \cite{Chomaz_PRE} also  gives an
indication of a phase change in the same excitation energy region.

The caloric curve shows a monotonic increase in temperature and no
plateau region is apparent, in contrast to caloric curves seen for
heavier systems \cite{JBN0}. At the apparent critical excitation
energy the temperature is $8.3 \pm 0.5$ MeV. Taking this to be the
critical temperature for this system,  we extracted the critical
exponents $\beta$, $\gamma$ and $\sigma$. The deduced values are
consistent with the values of the liquid gas phase transition
universality class \cite{Stanley}.

Since some fluctuation observables, such as the structure of the
Campi plot, $Z_{max}$ fluctuations etc., could be produced by mass
conservation effects where no assumption of the critical behavior
is needed,  these observables by themselves do not guarantee that
the critical point has been  reached. What differentiates the
present work from previous identifications of points of critical
behavior in nuclei, in addition to the fact that these are the
lightest nuclei for which a detailed experimental analysis has
been made, is the comportment of the caloric curve and the
critical exponent extraction.  Taken together, this body of
evidence suggests a liquid gas phase change in an equilibrated
system at, or extremely close to,  the critical point. Detailed
theoretical confrontations with  models which include or exclude a
liquid gas phase transition are certainly interesting and welcome.
Some work along this line is in progress \cite{Ma_comp,Botivna}.

\acknowledgements

This work was supported by the U.S. Department of Energy and the
Robert A. Welch Foundation under Grant No. A330. The work of YGM
is also partially supported by the NSFC under Grant No. 19725521
and the Major State Basic Research Development Program in China
under Contract No. G2000077404. YGM is also grateful to Texas
A$\&$M University - Cyclotron Institute for support and
hospitality extended to him. R.A., A.M-D and A. M-R wish to
acknowledge partial support from DGAPA and CONACYT-Mexico.

{}

\end{document}